\documentclass[12pt]{article}

\usepackage[english]{babel}
\usepackage[T1]{fontenc}
\usepackage{times}

\usepackage[a4paper,top=2cm,bottom=2cm,left=2cm,right=2cm,marginparwidth=1.75cm]{geometry}

\usepackage{amsmath, amsthm, amssymb, nomencl, mathtools, algorithmic, csquotes, subcaption, float}
\newcommand{\llbracket}{[\![}
\newcommand{\rrbracket}{]\!]}

\usepackage[ruled]{algorithm2e}
\usepackage{graphicx}
\usepackage{natbib}
\DeclareMathAlphabet{\mathmybb}{U}{bbold}{m}{n}

\usepackage{xurl}  
\usepackage[colorlinks=true, allcolors=blue]{hyperref}

\usepackage[capitalise]{cleveref}

\crefname{algorithm}{Algorithm}{Algorithms}
\crefname{figure}{Figure}{Figures}
\usepackage{comment}

\usepackage{tikz}
\usetikzlibrary{calc,trees,positioning,arrows,chains,shapes,decorations.pathmorphing,decorations.markings}

\usepackage{appendix}
\usepackage{tocloft}
\usepackage{etoc}
\usepackage{multirow}
\usepackage{longtable}

\usepackage{endnotes}
\let\footnote=\endnote

\usepackage{setspace}

\DeclareMathOperator*{\argmin}{arg\,min}

\allowdisplaybreaks

\newcommand{\bX}{\boldsymbol{X}}
\newcommand{\bx}{\boldsymbol{x}}
\newcommand{\bY}{\boldsymbol{Y}}
\newcommand{\by}{\boldsymbol{y}}

\newcommand{\bh}{\boldsymbol{h}}

\newcommand{\bz}{\boldsymbol{z}}

\newcommand{\calL}{\mathcal{L}}
\newcommand{\calC}{\mathcal{C}}
\newcommand{\calZ}{\mathcal{Z}}
\newcommand{\calX}{\mathcal{X}}
\newcommand{\calY}{\mathcal{Y}}

\newcommand{\reals}{\mathbb{R}}
\newcommand{\expect}{\mathbb{E}}
\newcommand{\proba}{\mathbb{P}}
\newcommand{\loss}{L}
\newcommand{\given}{|}

\newcommand{\round}[1]{\ensuremath{\lfloor#1\rceil}}

\title{Multilingual hierarchical classification of job advertisements for job vacancy statistics}

\author{
Maciej Beręsewicz\footnote{Poznań University of Economics and Business, Institute of Informatics and Quantitative Economics, Department of Statistics, Al. Niepodległości 10, 61-875 Poznań, Poland, E-mail: \url{maciej.beresewicz@ue.poznan.pl}; Statistical Office in Poznań, ul. Wojska Polskiego 27/29 60-624 Poznań, Poland.},
Marek Wydmuch\footnote{Poznan University of Technology, Institute of Computing Science, Piotrowo 2, 60–965, Poznań, Poland, E-mail: \url{mwydmuch@cs.put.poznan.pl}},
Herman Cherniaiev\footnote{University of Information Technology and Management in Rzeszów, and Educational Research Institute -- National Research Institute in Warsaw, Poland}, 
Robert Pater\footnote{University of Information Technology and Management in Rzeszów, and Educational Research Institute -- National Research Institute in Warsaw, Poland}
}
\date{}

\begin{document}
\maketitle

\doublespacing

\begin{abstract}

The goal of this paper is to develop a~multilingual classifier and conditional probability estimator of occupation codes for online job advertisements in accordance with the International Standard Classification of Occupations (ISCO) extended with the Polish Classification of Occupations and Specializations (KZiS), which is analogous to the European Classification of Occupations. In this paper, we utilise a~range of data sources, including a~novel one, namely the Central Job Offers Database, which is a~register of all vacancies submitted to Public Employment Offices. Their staff members code the vacancies according to the ISCO and KZiS. A~hierarchical multi-class classifier has been developed based on the transformer architecture. The classifier begins by encoding the jobs found in advertisements to the widest 1-digit occupational group, and then narrows the assignment to a~6-digit occupation code. We show that incorporation of the hierarchical structure of occupations improves prediction accuracy by 1-2 percentage points, particularly for the hand-coded online job advertisements. Finally, a~bilingual (Polish and English) and multilingual (24 languages) model is developed based on data translated using closed and open-source software. The open-source software is provided for the benefit of the official statistics community, with a~particular focus on international comparability.
\end{abstract}

Keywords: hierarchical multi-class classification; ISCO; ESCO; occupation; skills demand

\clearpage

\doublespacing

\section{Introduction}

Online and administrative sources yield substantial data rich in pertinent information that can be leveraged to enhance the understanding of labour market dynamics, as demonstrated by previous research \citep{hersh2018big, beresewicz2021}. In particular, data from online job advertisements (OJA) have been employed extensively in the tracking of changes in vacancies (see, for example, \cite{berkesewicz2021inferring, turrell2021uk, hershbein2018recessions}). The analysis of such information allows for the examination of the impact of current global shocks on labour demand. While the estimation of the number of jobs is a~crucial aspect, the analysis and forecasting of structural change represents a~significant challenge in economics during labour market transformation. The analysis of skill demand necessitates a~disaggregate approach, as the general definition of skills is unable to capture the specificities in firms' demand for skills.

Lightcast (\url{https://lightcast.io/about/data}), the largest global company providing statistics on OJA, classifies occupations but does not disclose the models or the quality of said models (e.g., classification errors). The European Union agency, Cedefop, is responsible for the collection of OJA data across all EU member states, which is then made available for analysis through the Skills-OVATE tool\footnote{See \url{https://www.cedefop.europa.eu/en/tools/skills-online-vacancies/occupations/}}. Cedefop classification of occupations is based on the International Standard Classification of Occupations (ISCO) up to the unit group level. Initially, they employed Support Vector Machines (SVMs), which Cedefop identified as the optimal approach in terms of classification accuracy \citep[cf. ][]{boselli2017using,colombo2018applying}.

A more recent machine learning approach dedicated to job offers was proposed by \citet{turrell20226}. The researchers employed a~corpus of 15.2 million advertisements posted between 2008 and 2016 on \url{Reed.co.uk} for the purpose of classifying a~job offer according to the 3-digit codes of the UK Office for National Statistics' (ONS) Standard Occupational Classification (90 minor occupational groups). In order to achieve this, the researchers employed a~fuzzy matching technique. The researchers report that manual assignment of occupations yielded 76\% correctness (on a~sample of 330 job offers), while their algorithm achieved 91\% accuracy (on a~sample of 67,900 job offers). However, the authors also indicate that their assessment was based on only 34\% of the initially collected job offers due to the inability of the ONS algorithm to produce a~confident label. In their study, \citet{ml-esco} employed a~pre-trained multilingual XLM-RoBERTa \citep{conneau2020unsupervised} transformer model (a~type of deep neural network; see, for example, \cite{vaswani2017attention}) to map national occupational classifications to the European Skills, Competences, Qualifications and Occupations (ESCO) framework. In addition to national qualifications and the ESCO dictionary, the authors make use of descriptions drawn from the Qualification Dataset Register and European Employment Services (EURES) online job advertisements. The authors conclude that the model demonstrates promising results when applied to work in a~variety of languages. \citet{chen2024explainable} put forth the utilisation of explainable AI (XAI) methodologies to develop a~lexicon-based approach. The method was applied to identify traditional and green energy jobs in online advertisements, with a~reported precision rate of 82\%. In 2024, Eurostat, in collaboration with Cedefop, launched a~text classification competition at the ISCO level. They provided a~hand-coded dataset comprising 26,000 OJA, with the highest score for the lowest common ancestor at 58\% (for further details, see \url{https://statistics-awards.eu/}). It should be noted that neither the data nor the algorithms are disclosed, so they cannot be used by researchers or the official statistics community.

In this study, we propose a~hierarchical conditional probability estimator and a~classifier of job advertisements for the purpose of estimating job vacancy statistics. The methodology is then applied to predict the probability and classify the advertisements across occupations according to the KZiS, which organises occupations into a~five-level hierarchy of job categories, represented by a~six-digit code. Each digit encodes the child category, with the exception of the final level, which is encoded by two digits. The 1-digit code represents the widest major category, while the 6-digit code corresponds to the fifth level, the most narrow category. This imposes a~classical hierarchical multi-class classification problem. 
Similarly to the approach taken in \citet{ml-esco}, our classifier is based on a~pre-trained transformer model, which we have carefully fine-tuned for the specific task of job classification.  The approach was tested using XLM-RoBERTa and HerBERT, a~transformed model pre-trained on a~large multilingual corpus that included the Polish language \citep{mroczkowski-etal-2021-herbert}. The classifier was trained and evaluated using a~dataset of online job advertisements gathered from Polish online job boards and administrative records.

Our contribution can be summarised as follows:

\begin{itemize}
    \item We propose and develop a~hierarchical conditional probability estimator and classifier that takes into account the structure of the official classification of occupations. This approach may serve as a~significant complement to traditional hand-coding techniques and facilitate the utilisation of big data sources for official statistics.
    \item We employ a~novel data source: large-scale administrative data comprising job offers that have been hand-coded by experts from public employment offices. In addition to the data itself, we provide a~detailed description of its quality, along with the experts' coding, which is often absent or unavailable in the existing literature.
    \item We developed a~model for 24 languages included in the European Union, with a~view to studying the performance of these languages.
    \item The open-source software and models are made available to the official statistics community so that they may benefit from the proposed methodology, either by utilising the classifiers or by modifying them to suit their own purposes.
\end{itemize}

The remaining part of the paper is as follows. \cref{sec-kzis} describes Polish Classification of Occupations and Specialities (hereinafter KZiS) and its relation to the International Standard Classification of Occupations (ISCO). \cref{sec-models} provides theory for the proposed approach that takes into account the hierarchical structure of official classifications. \cref{sec-data} describes the dataset used for training and testing, in particular administrative data sources, hand-coded online job offers and other available sources. \cref{sec-quality} contains information about the quality of the data, in particular how accurate are the experts' coding. \cref{sec-setup} presents experimental setup and its implementation. \cref{sec-results-classification} presents results for the proposed classifiers for Polish, bilingual (Polish and English) and multilingual datasets. The paper concludes with a~discussion and an exploration of the potential for extending the approach to other classifications, including the Occupational Information Network (O*NET), the European Skills, Competences, Qualifications and Occupations (ESCO) framework, and the Standard Occupational Classification (SOC). The supplementary materials provide a~brief tutorial on the use of the software, as well as detailed results. 

\section{Classification of occupations and specializations}\label{sec-kzis}

The KZiS classification is created by the Ministry of Family, Labour and Social Policy and provides basic information about job type \citep{kzis2014}. It~is based on ISCO established by the International Labour Organization (ILO). The current classification, available since 2023 is based on ISCO-08 and updated with new occupations, especially from the vocational education system, that are annually announced by the Ministry of National Education\footnote{\url{https://www.gov.pl/web/edukacja/zawody-szkolnictwa-branzowego} (as of 2024.07.31).}.

The KZiS is a 6-digit, hierarchically structured classification system. The hierarchy encompasses a range of occupational groups, from one-digit major groups to six-digit occupations (cf. \cref{tab-structure}). It develops the ISCO classification, which ends with 4-digit codes (unit groups). In this case, it is similar to the later established European Skills, Competences, Qualifications and Occupations (ESCO) classification. The KZiS classification encompasses over 2,500 occupations, which are grouped into ten major categories. In addition to the Armed Forces (group 0), the classification is primarily delineated by the level of job position (or skill) through a process of alignment with the UNESCO International Standard Classification of Education (ISCED 2013) and the levels of the Polish Qualifications Framework, as applicable to the majority of occupations within a~given group.

\begin{table}[ht]
\centering
\caption{Structure of the classification of occupations and specializations (as of 01.01.2023)}
\label{tab-structure}
\resizebox{\linewidth}{!}{
\begin{tabular}{lrrrrr}
  \hline
(1 digit) &  2 digits & 3 digits & 4 digits & \multicolumn{2}{c}{6 digits}  \\ 
Major Group & Sub-major & Minor &  Unit & with & without  \\ 
  & Groups & Groups & Groups & the rest & the rest  \\ 
  \hline
  0 -- Armed Forces Occupations &   3 &   3 &   3 &   3 &   3 \\ 
  1 -- Public Authorities, Senior Officials and Managers &   4 &  11 &  31 & 202 & 172 \\ 
  2 -- Professionals &   6 &  31 &  99 & 789 & 705 \\ 
  3 -- Technicians and Associate Professionals &   5 &  20 &  87 & 610 & 535 \\ 
  4 -- Clerical Support Workers &   4 &   8 &  27 &  89 &  71 \\ 
  5 -- Services and Sales Workers &   4 &  13 &  39 & 166 & 137 \\ 
  6 -- Skilled Agricultural, Forestry and Fishery Workers &   3 &   9 &  17 &  63 &  54 \\ 
  7 -- Craft and Related Trades Workers &   5 &  14 &  69 & 476 & 412 \\ 
  8 -- Plant and Machine Operators and Assemblers &   3 &  14 &  41 & 387 & 349 \\ 
  9 -- Elementary Occupations &   6 &  11 &  32 & 126 & 111 \\ 
  \hline
  All categories & 43 & 134 & 445 & 2,911 & 2,549 \\ 
   \hline
\end{tabular}
}
\begin{flushleft}
\begin{scriptsize}
Note: The term "the rest" is used to denote those occupations that are not classified elsewhere. To illustrate, the 4-digit occupation code "2522 Systems Administrators" is extended with two specific occupations: "252201 Computer Systems Administrator" and "252202 Integrated Management Systems Administrator". Additionally, the third category is "252290 Other Computer Systems Administrators".
\end{scriptsize}
\end{flushleft}
\end{table}

The majority of occupations within major group 1 and 2 require individuals to have completed a tertiary education qualification. These roles are associated with the highest skill levels, which entail the worker engaging in intricate practical and technical tasks (skill level 3), problem-solving, decision-making, and creativity (skill level 4). The majority of occupations within Groups 3 to 5 typically necessitate upper secondary education, frequently in a vocational context. These occupations are associated with ISCO skill levels 2 and 3. Skill level 2 encompasses the ability to perform tasks such as operating machinery, mechanical and electronic equipment, vehicles, and the manipulation, ordering, and storage of information. In order to perform tasks associated with ISCO skill level 2, secondary education is required for occupations within groups 6 to 8. Positions within major group 9 typically require a secondary education at most. Workers in these roles typically engage with tasks at levels 1-3 of the ISCO skill spectrum. Level 1 is associated with the completion of routine and elementary physical and manual tasks. The following section presents an overview of our methodology, which incorporates a consideration of the hierarchical structure of the occupation classification.

\section{Hierarchical multi-class classification}\label{sec-models}

Hierarchical multi-class classification is a~classical machine learning problem in which an object can be assigned a~set of classes that are hierarchically organized in the form of a~tree~\citep{sun2001hierarchical,silla2011survey}, and a~maximum of one class can be predicted per hierarchy level. 
Additionally, it is often required that every prediction must be coherent with respect to the hierarchy, meaning that for every predicted class, the classifier must also predict all its ancestors in the hierarchy. In other words, hierarchical multi-class classification can be seen as a~problem of selecting \textit{a~single path from the tree root to a~leaf}.
Because we are interested in population analysis of the job market, we not only require the classifier to assign discreet classes but also provide probability estimates of an online job offer belonging to the specific class at each level of the hierarchy.

\subsection{Formal setting}

Let $\calC = \{1, \ldots, m\}$ be a~finite set of $m$ classes with hierarchical relations between them encoded using a~prefix code.
Any such code can be given in the form of a~hierarchy tree in which a~path from a~tree root to a~leaf node corresponds to a~code word.
Under the coding, each class $c \in \calC$ is uniquely represented by a~tree node,
that is identifiable by a~code $\bz^l(c) = (z_1(c), \ldots, z_l(c)) \in \calZ^l$, where $l$ is the length of the prefix and $\calZ^l$ is a~set of all possible codes of length $l$. 
Only the tree root has no class assigned to it, and its code is of length 0, $\bz^0 = \varnothing$. 
The class $c$ with the code $\bz^l(c)$ of length $l$ is an ancestor of a~class $c'$ with the code $\bz^{l'}(c')$ of length $l' > l$ if and only if $\forall_{j=1}^l z_j(c) = z_j(c')$. If $l' = l + 1$, the class $c$ is called a~parent class of $c'$ and $c'$ a~child class of $c$.

For $z_i(c) \in \{0,1\}$, the code and the hierarchy tree are binary. In general, the code alphabet can contain any number of symbols.
Furthermore, $z_i$ can take values from different sets of symbols depending on the previous values in the code. In other words, the code can result in nodes of a~different parity even in the same tree. 
While this is not a~case considered in this work, the tree does not need to be perfectly balanced, resulting in code words of different lengths for different leaf nodes.
We present an example of a~hierarchy tree in \Cref{fig:class-hierarchy}.

\begin{figure}[ht!]
    \centering
\begin{tikzpicture}[scale = 1,every node/.style={scale=1},
	regnode/.style={circle,draw,minimum width=1.5ex,inner sep=0pt},
	leaf/.style={circle,fill=black,draw,minimum width=1.5ex,inner sep=7pt},
	pleaf/.style={rectangle,rounded corners=.5ex,draw,font=\small,inner sep=7pt, align=center}, 
	pnode/.style={rectangle,rounded corners=.5ex,draw,font=\small,inner sep=7pt, align=center},
	rootnode/.style={rectangle,rounded corners=.5ex,draw,font=\small,inner sep=8pt},
	level 1/.style={sibling distance=20em, level distance=10ex},
	level 2/.style={sibling distance=6em, level distance=15ex},
	]
	\node (z) [rootnode] {$\bz^0 = \varnothing$}
				child {node (a) [pnode] {$c = 1$\\$\bz^1 = (0)$} 
					child {node (b) [pleaf] {$c = 3$ \\ $\bz^2 =(0,0)$} edge from parent node[above left]{$1$}}
     				child {node (c) [pleaf] {$c = 4$ \\ $\bz^2 =(0,1)$} edge from parent node[above left]{$2$}}
					child {node (d) [pleaf] {$c = 5$ \\ $\bz^2 =(0,2)$} edge from parent node[above right]{$3$}}
					edge from parent node[above left] {$1$}
				}
				child {node (j) [pnode] {$c = 2$ \\ $\bz^1 = (1)$}
					child {node (k) [pleaf] {$c = 6$ \\ $\bz^2 =(1,0)$} edge from parent node[above left]{$1$}}
					child {node (l) [pleaf] {$c = 7$ \\ $\bz^2 =(1,1)$} edge from parent node[above left]{$2$}}
                    child {node (m) [pleaf] {$c = 8$ \\ $\bz^2 =(1,2)$} edge from parent node[above right]{$3$}}
                    edge from parent node[above right] {$2$}
			};
\end{tikzpicture}
\caption{An example of class hierarchy with 8 classes and their assigned codes $\bz$ organized into two levels.}
\label{fig:class-hierarchy}
\end{figure}

Let $\mathcal{X}$ denote an instance space, and an instance $\bx \in \calX$ be associated with a~single code word $\bz$ corresponding to a~leaf node in the hierarchy tree,
which corresponds to a~subset of classes $\calC_{\bx}$. being assigned to the $\bx$. 
This subset is often called a~set of relevant labels, while the compliment $\calL \backslash \calL_{\bx}$ is considered as irrelevant for $\bx$.
We identify a~set $\calL_{\bx}$ of relevant labels with a~binary (sparse)
vector $\by = (y_1, \ldots, y_m)$, in which $y_c = 1 \Leftrightarrow c \in \calL_{\bx}$.
By $\calY = \{0, 1\}^m$ we denote a~set of all possible label vectors.
We assume that observations $(\bx, \by)$ are generated independently and identically according to the
probability distribution $\proba(\bX = \bx,\bY = \by)$ (denoted later by $\proba(\bx, \by)$) defined on $\calX \times \calY$).

Our goal is to find a~\emph{classifier} $\bh(\bx) = (h_1(\bx),\ldots, h_m(\bx))$, 
which in general can be defined as a~mapping $\calX \rightarrow \reals^m$, that minimizes the \emph{expected loss} (or \emph{risk}):  
\begin{equation}
\loss_\ell(\bh) \coloneqq \expect_{(\bx,\by) \sim \proba(\bx,\by)} \left[ \ell(\by, \bh(\bx)) \right]\,,
\end{equation}
where $\ell(\by, \hat{\by})$ is the  (\emph{task}) \emph{loss}.

In particular, as our goal is to use the classifier for population analysis, we aim at estimating the conditional probabilities of classes $\eta_c(\bx) \coloneqq \proba(y_c = 1 \given \bx) = \proba(\bz^l = \bz^l(c) \given \bx)$. 
To obtain the estimates of conditional probabilities, one can use the class-wise log loss as a~surrogate objective:
\begin{equation}
\ell_{\log}(\by, \bh(\bx)) = \sum_{c=1}^m \ell_{\log}(y_c, h_c(\bx)) = \sum_{c=1}^m  \left ( y_c \log(h_c(\bx)) + (1-y_c) \log(1-h_c(\bx)) \right) \,.
\label{eq:log-loss}
\end{equation}
Then the expected class-wise log loss for a~single $\bx$ (i.e., the so-called \emph{conditional risk}) is:
\begin{equation}
\expect_{\by \sim \proba(\by \given \bx)} \left[ \ell_{\log}(\by, \bh(\bx)) \right] = \sum_{c=1}^m \expect_{\by \sim \proba(\by \given \bx)} \left[ \ell_{\log}(y_c, h_c(\bx)) \right] = \sum_{c=1}^m \loss_{\log}(h_c(\bx) \given \bx)\,.
\end{equation}
Therefore, it is easy to see that the point-wise optimal prediction for the $c$-th class is given by:
\begin{equation}
h_c^*(\bx)  = \argmin_h \loss_{\log}(h_c(\bx)\given \bx) = \eta_c(\bx) \,.
\end{equation}
Let us notice that under class hierarchy and the assumption that only one leaf node is assigned to instance $\bx$,
the conditional probabilities of labels obey the following relations:
\begin{itemize}
    \item The conditional probability of any parent class is equal to a~sum of conditional probabilities of its children:
    \begin{equation}
    \proba(\bz^l | \bx)  = \sum_{z_{l+1}} \proba\left(\bz^l \cup (z_{l+1}) | \bx\right) \,.
    \label{eq:parent-proba-sum}
    \end{equation}
    \item Since only a~single class per hierarchy level can be relevant, the sum of all conditional probabilities of classes on the same hierarchy level (with the same length of the codes $l$) should sum to 1:
    \begin{equation}
    \sum_{\bz^l \in \calZ^l} \proba \left( \bz^l | \bx \right) = 1 \,.
    \label{eq:level-proba-sum}
    \end{equation}
\end{itemize}

\subsection{Modeling probabilities under class hierarchy}
\label{sec:modeling-class-hierarchy}

We are not only interested in possible accurate estimates of conditional probabilities $\eta_c(\bx)$ for each class $c$, we additionally require for the predictions of the classifier to be coherent under class hierarchy, that is, obey the relations presented in \Cref{eq:level-proba-sum,eq:parent-proba-sum}.
Naively training the classifier using class-wise log loss \labelcref{eq:log-loss} does not guarantee that class relations will hold.
Because of that, in this work, we consider two classical approaches that guarantee coherency of prediction under class hierarchy.

We refer to the first approach as a~\emph{``bottom-up''}~\citep[cf.][]{barbedo2006automatic}. 
The bottom-up approach is one of the simplest to deal with the hierarchical classification problem. 
In this type of classifier, the conditional probability estimator of classes is trained only for classes in the leaf nodes of the hierarchy tree (i.e., 6-digit KZiS codes), and information about the class hierarchy is ignored. 
Because, out of these classes, only one can be positive at for instance, one can use categorical cross-entropy loss (a generalization of log loss) to train the probability estimator:
\begin{equation}
\ell_{\text{cce}}(\by, \bh(\bx)) = \sum_{c \in \calC'} y_c \log(h_c(\bx)) \,,
\label{eq:cce}
\end{equation}
where $\calC'$ is some subset of all the classes $\calC$, in this case being a subset of classes without any children.
During inference, after predicting probabilities for leaf labels, the predictions for labels higher in the hierarchy are reconstructed by summing the estimates of their direct children as in \labelcref{eq:parent-proba-sum} until the root node is reached.

We refer to the second approach we consider as the \emph{``top-down''} approach. 
In this algorithm, the probability of a~given class is
determined by a~sequence of decisions made by node classifiers that predict subsequent values of the
code word. By using the chain rule of probability, we~obtain:
\begin{equation}
    \eta_c(\bx) = \proba \left(\bz^l(c) \given \bx \right) = \prod_{i=1}^l \proba \left(z_i(c) \given \bz^{i-1}(c), \bx \right).
\end{equation}
Let us notice that since we deal here with a~multi-class distribution, we have that:
\begin{equation}
\sum_{z_i} \proba\left(z_i \mid \boldsymbol{z}^{i-1}, \boldsymbol{x}\right)=1    
\end{equation}
The classifier is then formed by the training probability estimator for every $\proba(z_i(c) \given \bz^{i-1}(c), \bx)$. This can be done using the cross-entropy loss \labelcref{eq:cce} introduced earlier. 
This approach is known in the literature as the nested dichotomies~\citep{Fox_1997} in statistics, hierarchical softmax (HSM) approach~\citep{Morin_Bengio_2005}, conditional probability estimation trees~\citep{Beygelzimer_et_al_2009_CPT} or probabilistic classifier trees~\citep{Dembczynski_et_al_2016}


\subsection{Statistical challenge of long-tail distribution}

It is common in classification tasks with a~large number of labels that labels are highly unbalanced -- only a~few classes have a~large number of data points, while the vast majority is represented by a~much smaller number of examples. This kind of distribution is said to be \emph{long-tailed}~\citep{Bhatia_et_al_2015, Babbar_Scholkopf_2017_Dismec}.
The long-tail distribution imposes statistical challenges in modeling the conditional probability distribution using classical machine learning algorithms like logistic regression or naive Bayes classifier, as they are not performing well under such a~low data regime. 

\begin{figure}[ht!]
    \centering
    \includegraphics[width=0.8\textwidth]{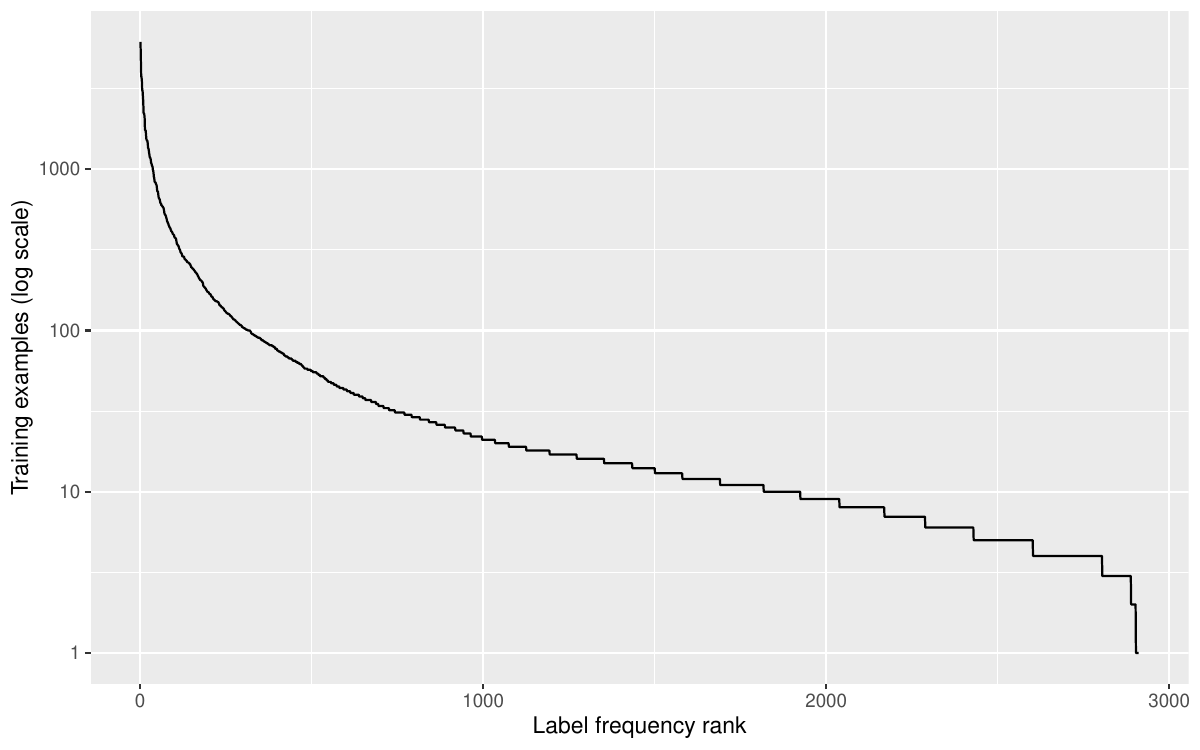}
    \caption{
    Last level classes (codes) frequency in the study datasets. 
    The X-axis shows the class rank when sorted by the frequency of positive instances, and the Y-axis shows the number of positive instances.}
    \label{fig-frequency}
\end{figure}

In our task, the classifier must predict a~set of classes out of 2,911 possible ones. \cref{fig-frequency} illustrates the long-tail distribution present in the dataset, which is described in detail in \cref{sec-data}. It is evident that almost one-third of the classes are represented by less than 10 examples, while only around 250 classes have more than 100.

\subsection{Proposed model architecture}
\label{sec:proposed-architecture}

To address long-tail distribution, instead of training the classifier from scratch, we opted for using a~pre-trained transformer model. We are focusing on the popular family of models based on the Bidirectional Encoder Representations from Transformers (BERT;~\cite{devlin2018bert}) architecture. 

BERT is a~pre-trained language model that has revolutionized natural language processing (NLP). It is an encoder-only model, meaning it is designed to create representations of an input text in the form of a~vector that can be used for a~variety of NLP tasks such as sentiment analysis, spam detection, or any other text classification task. BERT model uses bidirectional Transformer architecture. This mechanism helps the model weigh the importance of each word in the input sequence by considering its relationship with all other words. It allows BERT to capture context in a~bidirectional way, meaning it considers both the left and right context when processing words.

It can be pre-trained on vast amounts of data in an unsupervised manner and can be fine-tuned for downstream tasks such as considered text classification.
BERT models are usually pre-trained using two key strategies:
\begin{itemize}
    \item Masked Language Modeling (MLM): A portion of the input tokens are randomly masked, and the model is trained to predict these masked tokens based on their context.
    \item Next Sentence Prediction (NSP): This task trains the model to understand sentence relationships by predicting whether two sentences appear consecutively in the original text.
\end{itemize}

In this work, we take a~pre-trained BERT model and use its predicted text representation (hidden representation of [CLS] Token) combined with bottom-up or top-down output with linear models for modeling conditional probabilities of classes. We carefully fine-tune the resulting architecture in an end-to-end fashion -- both partners of bottom-up and top-down outputs are updated, as well as the parameters of the BERT transformer model. This way, we obtain a~model that is well-adjusted to the task while retaining its vast general language knowledge, that is, about words and concepts unseen in the training set used for fine-tuning. This should allow it to perform well also on a~long-tail -- classes with a~small number of training examples.  In the next section we describe the data used to train and test the proposed models.

\section{Data sources}\label{sec-data}

\subsection{Official classifications and definitions}

This dataset has been prepared for classification using the official classifications and accompanying documentation available on the website of the Ministry of the Family, Labour and Social Policy.

\begin{figure}[ht!]
    \centering
    \includegraphics[width=0.48\linewidth]{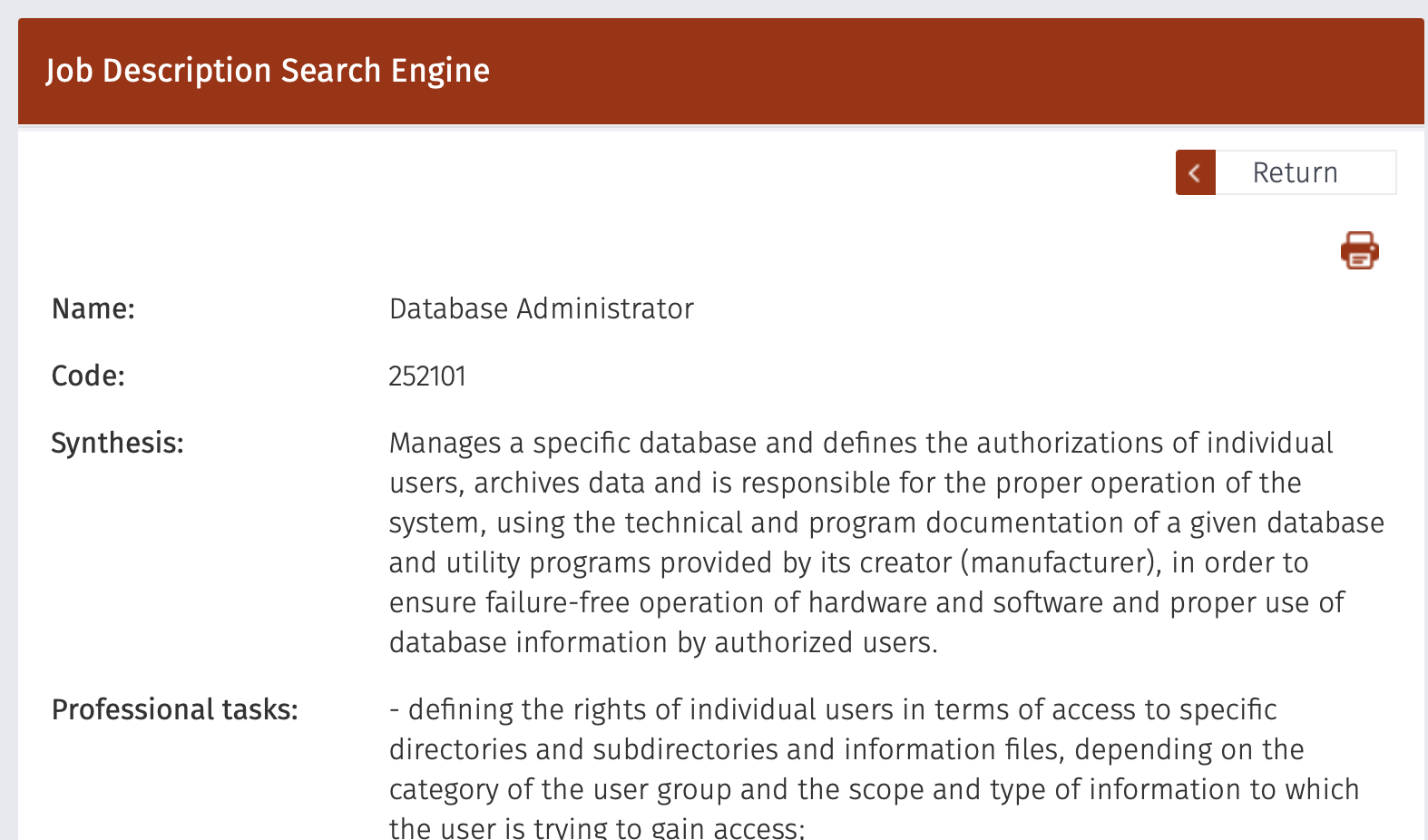}
    \includegraphics[width=0.48\linewidth]{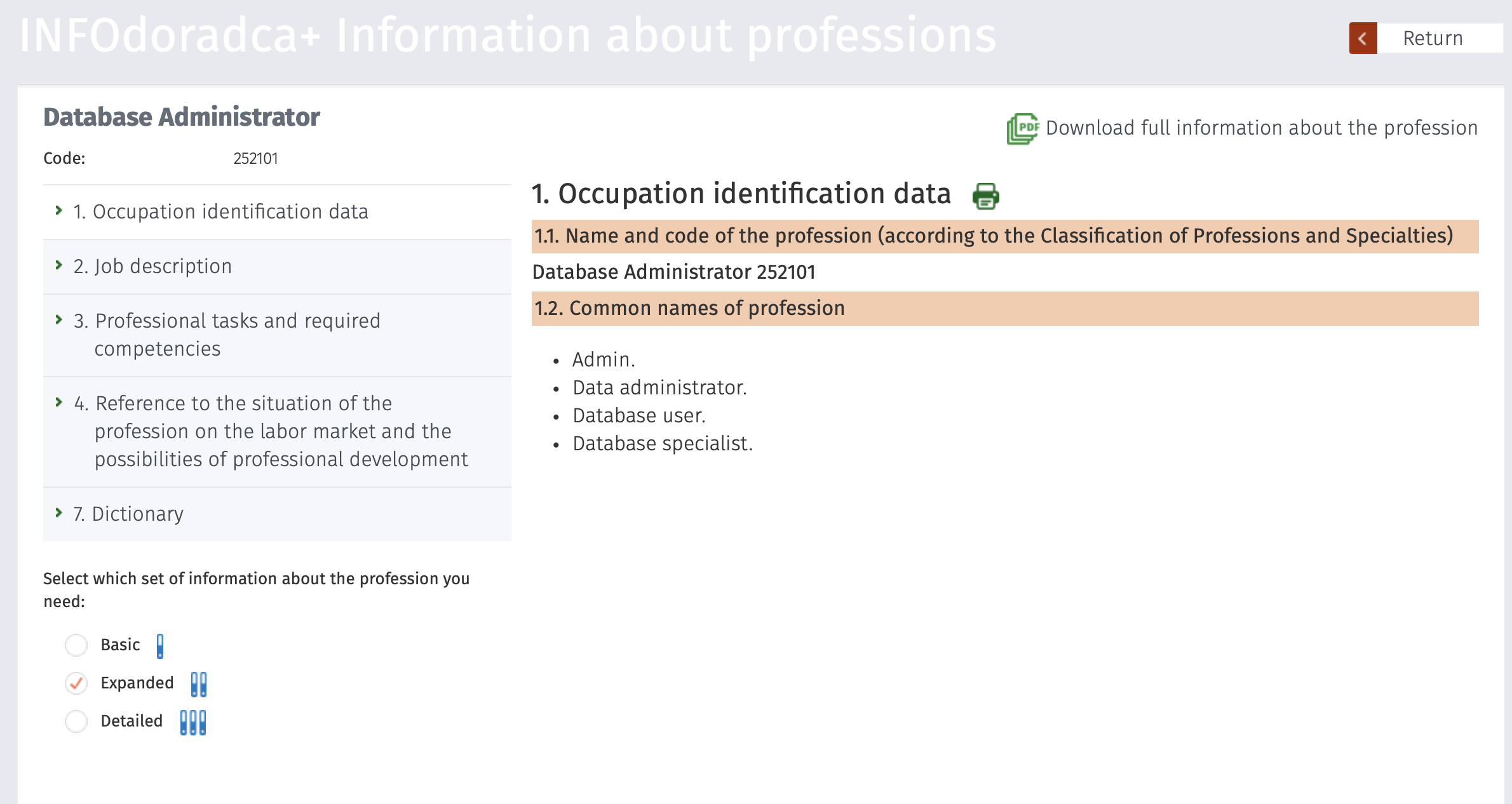}
    \caption{Example screenshots of the job description engine and INFOdoradca webservice}
    \label{fig-screenshots}
\end{figure}

\begin{itemize}
    \item Job description search engine\footnote{Available at \url{https://psz.praca.gov.pl/rynek-pracy/bazy-danych/klasyfikacja-zawodow-i-specjalnosci/wyszukiwarka-opisow-zawodow}  (as of 2024.07.18).} -- This service provides a comprehensive description of all occupations, including the name, code, synthesis, and a list of elementary and additional job tasks. For illustrative purposes, please refer to \cref{fig-screenshots}. It is important to note that the descriptions of the occupations exhibit considerable variation in terms of both detail and completeness. 
    \item INFOdoradca+ webservice\footnote{\url{https://psz.praca.gov.pl/rynek-pracy/bazy-danych/infodoradca} (as of 2024.07.18).} -- The service comprises a~curated list of 1,000 occupations, offering a~comprehensive overview of the profession or job role in question. It provides detailed information on the specific skills, qualifications and authorisations required for entry into the profession or job role, as well as the tasks and competencies necessary for success in the field. For an illustrative example, please refer to \cref{fig-screenshots}, which depicts a~description from the INFOdoradca+ web service.
\end{itemize}

In addition to the official classifications, the following datasets were employed:

\begin{itemize}
    \item thesaurus created by Statistics Poland\footnote{Available at \url{https://stat.gov.pl/Klasyfikacje/doc/kzs/slownik.html} (as of 2024.07.18)} -- contains 1,338 unique occupations and their synonyms, with the number of synonyms varying from 2 (equivalent to approximately 40\% of all occupations) to 82 (for one occupation 813101 "Aparatowy procesów chemicznych" (Operator of chemical processes equipment)) with mean and median equal to 4.7 and 3 respectively.
    \item KZiS-ESCO linkage database -- list of selected occupations from KZiS linked to ESCO proposed by  \citet{stkechly2023propozycja}. This dataset contains 557 unique codes along with its descriptions downloaded from the ESCO website\footnote{The value of this crosswalk is providing occupation descriptions from the ESCO classification and connecting occupations with skills, what has not been done in the KZiS. To connect KZiS with ESCO authors used an expert method supported with language modelling and semantic similarity based on Sentence-BERT model. They determined the relations for all KZiS occupations by matching existing ESCO occupations to KZiS. In linking both classifications, the authors used several types of semantic relations: exact match, broader match, narrower match, and close match. They excluded false friend (no) matches and weak matches. Available at \url{https://esco.ec.europa.eu/en/classification/occupation_main} (as of 2024.07.18)}. In our dataset, we only used exact matches (about 8.7\% of all KZiS codes).
    \item Civil service job offers database (KPRM) -- 2,941 job offers from the civil service (\url{https://nabory.kprm.gov.pl}) for 12 occupation codes\footnote{The following codes were identified 121101, 121904, 214921, 242208, 242211, 242213, 261103, 315202, 315209, 325504, 334102, 421402.} that meet the official definitions. This database was selected based on clerical review of matching job titles and descriptions.
\end{itemize}

For the official classifications (dictionary and INFOdoradca+), KZiS-ESCO and KPRM we have developed web-scraping algorithms that collected all available information on a~given occupation. 

In order to ensure the accuracy of online job advertisement classification, it is essential to utilise a comprehensive range of resources beyond official dictionaries or a limited number of advertisements from the Civil Service. Consequently, we have chosen to employ actual job postings from two primary sources: the Central Job Offers Database (hereinafter ePraca; previously known as CBOP), which encompasses all job advertisements submitted to Public Employment Offices (PEOs), and hand-coded job advertisements from multiple online sources. The subsequent subsections will delineate the characteristics of these datasets.

\subsection{Central Job Offers Database -- ePraca}

Public employment services in Poland include public employment offices (PEOs), which operate at LAU1  (\textit{Local Administrative Units}) level and are responsible for registering and managing unemployment. This means that there are two sources of data about employment in Poland: (i) the Labor Force Survey (LFS) and (ii) registered data collected by PEOs. The register maintained by PEOs contains detailed characteristics of unemployed persons and job offers. The number of registered unemployed may be higher than the LFS estimate because registration is connected with free health insurance and unemployment benefits. 

The structure of job offers from PEOs differs from the general population of job vacancies. There is an over-representation of jobs from companies that have an incentive to advertise their vacancies through public employment offices, for example in the case of refunded internships or publicly-subsidised workplaces for the disabled \citep{berkesewicz2021estimating}. Public entities, in particular, are more willing to publish job offers in PEOs because they are often obliged to do so by their own internal regulations. Finally, low-payed jobs are more often sent to PEOs because people with lower qualifications often rely on public institutions to help them find a~job. Better-paying jobs are more often advertised on job boards, in media or through private HR agencies, which charges fees for their services \citep{galecka2015ile, radzikowski-covid}.

\begin{figure}[ht!]
    \centering
    \includegraphics[width=0.9\linewidth]{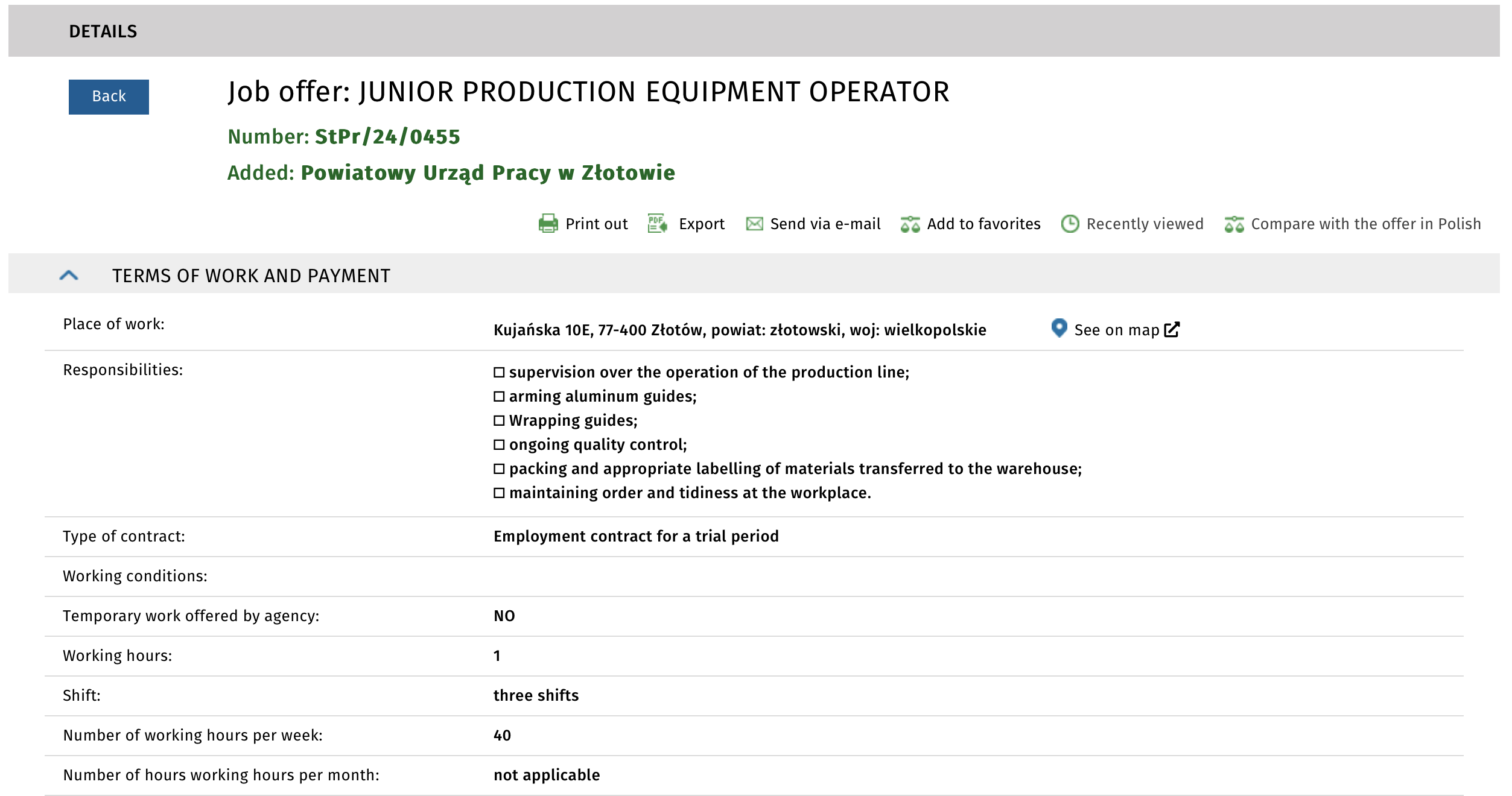}
    \caption{Example job posting from ePraca}
    \label{fig-cbop-example}
\end{figure}

Job offers may be submitted via a~paper or online form, following a~highly structured format\footnote{For an illustrative example, please refer to this website: \url{https://warszawa.praca.gov.pl/zgloszenie-oferty-pracy} (in Polish). In Supplementary Materials we provide an example paper form in Polish and English translated using DeepL software.}. The requisite information encompasses the entity's name, address, and REGON/NIP identifier; the job title; the number of vacancies; the type of contract; a~description of the responsibilities and requirements for the applicants. An example of a~job posting is provided in \cref{fig-cbop-example}. It is important to note that the option to specify the job occupation according to the official classification is available on the online form, though this is not a~mandatory requirement. In accordance with the findings of our discourse with PEOs, it has been established that the majority of employers refrain from populating this field, and instead, PEOs personnel undertake the coding of job advertisements according to the official KZiS classification manually. Regrettably, we are unable to present any numerical results to substantiate our claims, as this information is not accessible through ePraca database. In \cref{sec-quality}, we examine the quality of ePraca coding using a~sample of job ads. Information about the occupation is policy-relevant. The Ministry of National Education annually announces the forecast of the demand for employees in vocational education occupations on the national and regional labour markets\footnote{See \url{https://www.gov.pl/web/edukacja/prognoza-zapotrzebowania-na-pracownikow-w-zawodach-szkolnictwa-branzowego-na-krajowym-i-wojewodzkim-rynku-pracy-2024}}. Based on the forecast the Ministry differentiates the educational part of the general subsidy for vocational education students between local government units. The regional governments consider the forecast in preparing an annual list of occupations for which the costs of vocational training for young employees is reimbursed. The forecast is also used by regional labour market councils in issuing opinions on the validity of education in occupations.

The ePraca data is accessible via the API for registered users\footnote{More information is available at the following website \url{https://oferty.praca.gov.pl/portal/index.cbop\#/dlaInt}}. Each job advertisement comprises 174 fields in JSON format, including an occupation code according to the KZiS and descriptions provided by the employer or recruitment agency. The data used in this study was downloaded on a~daily basis from the beginning of 2022 to the end of 2023. The data was then subjected to a~process of cleaning, whereby erroneous codes or descriptions with missing data in all fields were removed\footnote{The codes used for this purpose are available for download online at \url{https://github.com/OJALAB/CBOP-datasets}}. Following this, the data set comprised over 822,000 fully labelled ads with 2,468 unique occupation codes and job ads descriptions, based on the aforementioned criteria. 

The distribution of KZiS codes was markedly uneven, with 145 codes having only a~single example, 128 codes having two examples, and codes with over 5,000 examples\footnote{The five most prevalent codes were: The following job titles were identified: 522301 "Sprzedawca" (Sales Assistant), 432103 "Magazynier" (Warehouse Operator), 931301 "Pomocniczy robotnik budowlany" (Construction Worker's Helper), 911207 "Pracownik utrzymania czystości" (Cleaner), 515303 "Robotnik gospodarczy" (Maintenance Worker).}. To reduce the size of the ePraca database and speed up training, a~stratified sample of the ePraca dataset was created, with strata defined by KZiS code and the number of characters in the description. A total of 11,572 strata were formed. \cref{tab-cbop-strata} presents information regarding the number of advertisements by the number of characters in the advertisement description. 

\begin{table}[ht!]
\centering
\caption{Number of ePraca ads by the number of characters in the description}
\label{tab-cbop-strata}
\begin{tabular}{lr}
  \hline
Number of characters & Number of ads/examples \\ 
  \hline
  $[0,50]$ & 6,628 \\ 
  $(50,100]$ & 50,576 \\ 
  $(100,200]$ & 196,240 \\ 
  $(200,300]$ & 172,942 \\ 
  $(300,400]$ & 118,598 \\ 
  $(400,500]$ & 79,699 \\ 
  $>500$ & 198,614 \\ 
   \hline
\end{tabular}
\end{table}

From each stratum, a~simple random sample was selected, with the size determined as $\max\left\{1, \round{0.2\times N_h}\right\}$, where $N_h$ is the strata sample size and $\round{}$ denotes rounding to the nearest integer. The final sample size was 167,244 (around 20\% of the ePraca dataset). 

\subsection{Hand coded datasets}

\subsubsection{The 10,000 ads dataset}

For our study, we have used two datasets of hand-coded job advertisements. The first consisted of 10,000 job advertisements randomly selected from a~large dataset of web-scraped advertisements. The second dataset consisted of 1,000 job advertisements selected from one of Poland's largest online recruitment services using specific keywords to match a range of occupations. We selected the IT sector for its diversity, particularly during the analysed period, and for jobs that are less prevalent in online recruitment portals. This study focuses on the 10,000 dataset.

As a~basis for the sample, we have used 1,805,967 job ads scrapped between 2021-06-28 and 2021-11-08 from 18 sources\footnote{Information about the number is presented in \cref{tab-app-sampling-frame} in Supplementary Materials.}. This dataset contained information on the web service (the source), url, language of title and description according to a~given website, job title, employer, region, job type, position level, contract type, salary, category (according to a~given source), web-scraping date and job description. Note that some columns contained missing data (e.g.,~salary).

Before sampling, the data was cleaned to remove rows that had no job description (233,464 observations), had zero words or less than 5 characters\footnote{The number of words was counted using \texttt{stri\_count\_words} function from the \texttt{stringi} package and number of characters were counted using \texttt{nchar} from the base R} (93 observations). Then we removed records with non-Polish descriptions. We have used Google's Compact Language Detector 3 available through the \texttt{cld3} package in R \citep{pkg-cld3} to detect the language of the job description and identified 181,425 rows with non-Polish text. In the end, the final dataset used for sampling consisted of 1,390,985 unlabelled records (77\% of the initial dataset).

Next, we have created two stratification variables. The first was based on the identification of specific occupations: "sprzedawca" or "kasjer" (seller or cashier), "doradca klient" (customer consultant), "magazynier" (warehouse worker), "przedstawiciel" (representative), "pracownik produkcji" (production worker), "pomoc kuchenna" or "kucharz" (kitchen help or chef), "biuro" (bureau), "sprzątacz" or "sprzatacz" (cleaner with and without Polish-specific letter) and "specjalista" (specialist). If none was matching "other" category was assigned. 

The second stratification variable was based on the number of words in the description that were categorized into the following groups: $[1, 10]$, $(10, 25]$, $(25, 50]$, $(50, 100]$, $(100, 150]$,  $(150, 200]$, $(200,300]$ and $300+$. \cref{tab-population-stratas} presents information about stratification and population sizes within these strata. The largest strata was "specjaliści" (specialists) with over 130k ads. The second largest were "sprzedawcy" (sellers) with close to 55k ads. The smallest group were "sprzątacze" (cleaners) with over 7k ads. According to the second stratification variable groups with over 50 words were more or less of the same size. There were about 40k ads with at most 10 words and 70k with at most 25 words. 

In order to code the dataset, three experts from PEOs were recruited in February 2022. Due to budgetary limitations, the sample size was set at 10,000, with each expert responsible for coding 3,333 advertisements. However, due to rounding, the final sample size was 10,002. The sample size was selected using a stratified systematic sampling design, with probabilities proportional to the number of characters in a given ad. The sample was selected using \texttt{strata} from the \texttt{sampling} package \citep{pkg-sampling}. Further details regarding the sample size according to the source can be found in \cref{tab-sample-size-source}. The primary source was pracuj.pl, praca.egospodarka and aplikuj.pl. It should be noted that 241 ads were from ePraca, which enabled an assessment of the quality of ePraca encoding. 

A total of 3,233 advertisements were selected for each expert, with an additional 100 advertisements sampled for cross-validation of expert coding\footnote{it should be noted that for 4 ads we also had information from ePraca}. Each expert was given an Excel spreadsheet, available via Google Sheets, containing all the relevant information. The experts were tasked with coding the advertisements into six-digit codes and providing comments where necessary. They were given a one-month period to complete the coding of the advertisements into occupation codes. 

\subsubsection{The 1,000 ads dataset}

In January 2023, an additional dataset of 1,000 advertisements was selected for manual coding by a single expert. The sample was selected from a comprehensive dataset of over 543,000 ads scraped from one of the largest services in Poland. In the case of this particular dataset, we were able to identify in excess of 227,000 unique descriptions. We then proceeded to implement a~sample selection process in accordance with the following scheme: the specification of a~search phrase (including regular expression notation) and the subsequent selection of the most appropriate samples. The information pertaining to the aforementioned phrases, the selected advertisements and the number of cases in one of the largest services in Poland is detailed in \cref{tab-1000-sample}.

The rationale behind the selection of the phases is as follows. Firstly, the decision was taken to include advertisements for positions related to computer science, data science and information technology. Secondly, the intention was to incorporate occupations that were not previously covered by our existing sources, such as pharmacist, sworn translator or piano repairer. 

\subsection{The multilingual dataset}

All datasets (over 200k rows) were translated into English and then to 22 other languages: Bulgarian, Czech, Danish, German, Greek, Spanish, Estonian, Finnish, French, Irish, Hungarian, Italian, Lithuanian, Latvian, Dutch, Portuguese, Romanian, Russian, Slovak, Slovenian, Swedish, Ukrainian. This procedure was done in two steps.

Firstly, the Polish text was translated into English using Google Translate. This was achieved by utilising the Google Sheets application and the \texttt{googletranslate()} function. This approach was employed for the translation of over 200,000 examples. Additionally, Argos Translate was employed for the translation of the remaining text. The open-source offline translation library, written in Python, is available at \url{https://github.com/argosopentech/argos-translate}. It uses the \texttt{OpenNMT} library \citep{openNMT}, which in turn uses the CTranslate2 library, which is a~fast inference engine for Transformer models. The objective was to evaluate the influence of an automated language library on the classification of a~bilingual dataset\footnote{Unfortunately, information on the quality of the automatic translation is rather limited. For instance, \citet{klein-etal-2020-opennmt} reports that the OpenNMT/CTranslate2 library scored 26.7\% in the Bilingual Evaluation Understudy (BLEU) task (see \citet{papineni2002bleu}) for English to German translation. Author of the Argos Translate fine-tuned the OpenNMT model in the OPUS -- A Collection of Multilingual Parallel Corpora with Tools and Interfaces (see \citet{OPUS}). Certainly, the quality of automated translation has an effect on the classifier, as we show in the comparison between the Polish-English translations using Google Translate and Argos Translate.}. 

Secondly, the ads translated to English with Google Translate descriptions were translated into 22 languages using the Argos Translate software. The principal rationale for this approach was the scarcity of human resources (there were no assistants available to undertake the translation process using Google Sheets) and time constraints, given that access to GPUs was available, as supported by the Argos Translate open-source library.

Due to budgetary constraints, an assessment of the quality of automated translation was not conducted. However, based on a~discussion with PEO experts, it was evident that if they were to code English or other language-based descriptions, they would likely utilise Google Translate or other free translation services.

\section{Quality of the input data}\label{sec-quality}

\subsection{Hand-coded data quality}

\subsubsection{The 1,000 dataset}

The 1,000 dataset was coded by a~single expert, and thus, cross-validation was not available. The expert identified 968 advertisements for which a~single occupation code was deemed appropriate, while 32 advertisements were assigned multiple codes (27 with two codes, three with three codes, and one with four codes). Therefore, the final dataset comprised 1,035 advertisements with 226 codes, with the most prevalent being: "Programista aplikacji" (Applications programmer, 125 occurrences; 251401), "Analityk baz danych (data scientist)" (Data scientist, 77 occurrences; 252102) and  "Specjalista do spraw cyberbezpieczeństwa" (Cybersecurity specialist, 63 occurrences; 252902).  See \cref{table-hand1k-strata} for the information on the number of unique 6-digit and 1-digit codes for a~given stratum defined by the search phrases.

\subsubsection{The 10,000 dataset}

The process of assessing the quality of the hand-coded data was an iterative one. Information about the coding is presented in \cref{tab-hand-coded-init-quality}. From the final sample, 32 advertisements with a~description in a~language other than Polish were removed, although they had been coded by experts. 

\begin{table}[ht!]
    \centering
    \caption{Information about the quality of hand-coded data}
    \label{tab-hand-coded-init-quality}
    \begin{tabular}{rrrrr}
    \hline
        Initial & Non-Polish  & Non-matching & Multiple & Final\\
        sample  &  description  & codes  & codes & sample \\
        \hline
        10,002 & 32 & 38 & 59 & 9,932\\
        \hline
    \end{tabular}
\end{table}

It was observed by experts that for 38 advertisements (0.38\%; 13 from the first expert, 13 from the second, and 2 from the third), it was not possible to ascertain the occupation due to a~lack of pertinent information or because no occupation in the official classification matched the description. In one instance, the expert indicated that the advertisement describes two distinct occupations. In 59 cases (representing 0.6\% of those with codes), experts noted that multiple codes align with the description. Rather than removing these 59 cases, we selected to flag them as potential codes for the same description. Consequently, the final dataset comprised 9,932 ads (prior to accounting for multiple codes).

The subsequent stage of the study involved an investigation into the quality of the coding produced by experts based on a~set of 100 ads that exhibited a~certain degree of overlap. Two of the experts did not provide a~code for one of the advertisements, whereas the others did. Accordingly, for the purposes of comparison, the total number of ads considered was 98. \cref{tab-consistency-experts} presents the data on the rate of agreement of expert coding at the 1-, 2-, 4-, and 6-digit levels (section "before clerical review") along with 95\% confidence interval using sampling weights using the \texttt{survey} \citep{pkg-survey} package. In addition, we have estimated Cohen's kappa coefficient for agreement between 1 digit codes for each pair of experts (we used \texttt{survey::svykappa()} function).

\begin{table}[ht]
\centering
\caption{Point and 95\% interval estimates of the rate of agreement of expert coding at 1, 2, 4, and 6 digits}
\label{tab-consistency-experts}
\resizebox{\linewidth}{!}{
\begin{tabular}{rrrrrr}
  \hline
Experts & 1 Digit & 2 Digits & 4 Digits & 6 Digits & Kappa \\ 
\hline
\multicolumn{6}{c}{Before clerical review} \\ 
  \hline
   1 \& 2 & 87.7 (80.5, 93.4) & 84.2 (76.6, 90.5) & 78.7 (70.3, 86.1) & 74.2 (65.2, 82.2) & 85.4 (77.8, 93.0)\\ 
   1 \& 3 & 86.2 (78.0, 92.7) & 86.2 (78.0, 92.7) & 74.7 (65.2, 83.2) & 66.2 (56.1, 75.7) & 83.5 (75.1, 92.0)\\ 
   2 \& 3 & 85.9 (77.7, 92.5) & 84.5 (76.1, 91.3) & 75.6 (66.4, 83.8) & 68.5 (58.7, 77.6) & 83.1  (74.4, 91.8)\\ 
   All & 79.9 (70.9, 87.6) & 78.5 (69.4, 86.3) & 66.1 (56.3, 75.3) & 59.2 (49.1, 69.0) & -- \\ 
   \hline
\multicolumn{6}{c}{After clerical review} \\
  \hline
  1 \& 2 & 92.6 (86.2, 97.1) & 90.3 (83.4, 95.5) & 84.4 (76.4, 91.0) & 80.5 (72.0, 87.8) & 91.2 (84.8, 97.6) \\
  1 \& 3 & 90.2 (83.1, 95.5) & 90.2 (83.1, 95.5) & 84.7 (76.4, 91.5) & 77.5 (68.2, 85.6) & 88.3 (81.1, 95.6)\\
  2 \& 3 & 86.8 (79.0, 93.0) & 85.8 (77.8, 92.2) & 80.8 (72.0, 88.3) & 72.9 (63.3, 81.5) & 84.2 (76.1, 92.4)\\
  All & 84.8 (76.6, 91.5) & 83.8 (75.4, 90.7) & 75.6 (66.2, 83.9) & 68.5 (58.8, 77.5) & -- \\ 
  \hline
\end{tabular}
}
\end{table}

In general, pairs of experts agree at the level of 86\% at the first digit of occupation codes, with this figure decreasing to 73\% and approximately 67\% at six digits. The level of agreement among all experts is 80\% for the first digit and decreases to approximately 60\% at the sixth digit.  The findings were presented to the experts, who were invited to provide comments and suggest potential amendments. Expert 1 elected to alter 10 of their original codes, Expert 2 chose to modify 13 of their initial codes, and Expert 3 revised six codes (including the addition of two missing codes). The results of the clerical review are presented in Table 2 (section 'After clerical review'). As anticipated, the agreement rate increased, although some discrepancies between experts remained. The overall agreement on a~single digit increased to 85\%, while for six digits it reached 68\%. The primary source of discrepancy was the lack of relevant information about the education or skills.  For all pairs of experts, the Kappa coefficient for agreement of 1 digit codes shows strong agreement with levels close to 85\%.

Next, we focus on the assessment of the quality of ePraca coding, as in our sample, we had 236 ads from this source. We were able to link 213 unique ads to the ePraca using the hash identifier present in the ad's URL. \cref{tab-consistency-experts-cbop} presents information on the quality of ePraca coding according to the experts (point and 95\% CI for the rate of agreement) and Kappa coefficient for each expert and ePraca as well as all experts (treated as one group) and ePraca as another (the last row denoted as \textit{All}). Note that due to sparse data, all measures were calculated without the 1~and 6~occupation group for both ePraca and experts (3 obs were dropped).

\begin{table}[ht!]
\centering
\caption{Point and 95\% interval estimates of the rate of agreement of expert coding at 1, 2, 4 and 6 digits for ePraca ads}
\label{tab-consistency-experts-cbop}
\resizebox{\linewidth}{!}{
\begin{tabular}{rrrrrrr}
  \hline
Expert & Count & 1 Digit & 2 Digits & 4 Digits & 6 Digits & Kappa \\ 
  \hline
    1 &  75   & 78.2 (65.9, 88.3) & 74.9 (62.9, 85.2) & 63.2 (50.7, 74.9) & 56.3 (43.7, 68.5) & 73.0 (59.2, 86.8)\\ 
    2 &  66   & 79.0 (67.5, 88.5) & 76.4 (64.6, 86.3) & 72.2 (60.3, 82.7) & 65.6 (53.5, 76.7) & 74.3 (61.8, 86.8)\\ 
    3 &  69   & 81.0 (70.3, 89.7) & 80.0 (69.3, 89.0) & 71.6 (59.8, 82.1) & 62.7 (50.1, 74.5) & 77.6 (66.2, 88.9)\\ 
    All & 210 & 79.3 (73.0, 85.0) & 77.0 (70.6, 82.8) & 68.8 (62.0, 75.2) & 61.4 (54.4, 68.1) & 75.2 (67.9, 82.4) \\
   \hline
\end{tabular}
}
\end{table}

Subsequently, an assessment of discrepancies in coding was conducted by experts. Two experts consented to participate in the study and provided comments. The second expert determined that six of the 29 advertisements provided by ePraca were accurate, differing from the expert's initial assessment. Additionally, four of the 29 advertisements exhibited a~dual accuracy, with both codes being correct. Consequently, the expert's agreement rate at the 6-digit level would increase to 76\%, while at the 1-digit level, it would increase slightly, given that the disagreements are at more detailed levels. The third expert determined that in seven of the 30 advertisements, both codes are accurate. However, the expert highlighted that the lack of sufficient information precludes the ability to distinguish between the two. Consequently, for this expert, the agreement rate at the six-digit level would increase to 72\%, while at the one-digit level, a~slight increase is observed due to the disagreements occurring at more detailed levels. This also results in lower Kappa coefficients around 75\% in comparison to about 85\% presented in \cref{tab-consistency-experts} but they are not significantly different if we compare confidence intervals.

\cref{tab-comparison-codes-cbop-expert} presents a~comparison of agreements at the one-digit level. The figures in the table refer to the precision of the classification of ePraca ads. With the exception of the 4th group, the agreement rate for groups 2 to 9 is above 70\%. However, as there is only one example in the first group, we refrain from commenting on it. Overall, it can be observed that the ePraca codes are characterised by a~high level of confidence, with the main discrepancies occurring between the second and fourth groups.

\begin{table}[ht!]
\caption{Comparison of codes at the 1-digit groups between ePraca and experts coding}
\label{tab-comparison-codes-cbop-expert}
\centering
\begin{tabular}{rrrrrrrrrr|r}
  \hline
  \multicolumn{11}{c}{Hand-coded} \\
ePraca & 1 & 2 & 3 & 4 & 5 & 6 & 7 & 8 & 9 & N \\ 
  \hline
  1 & \textbf{0.0} & 0.0 & 100.0 & 0.0 & 0.0 & 0.0 & 0.0 & 0.0 & 0.0 & 1\\ 
  2 & 0.0 & \textbf{80.8} & 12.4 & 0.0 & 6.8 & 0.0 & 0.0 & 0.0 & 0.0 & 19\\ 
  3 & 0.0 & 23.0 & \textbf{65.8} & 7.0 & 0.0 & 0.0 & 4.2 & 0.0 & 0.0 & 20\\ 
  4 & 0.0 & 21.8 & 3.9 & \textbf{68.5} & 0.0 & 0.0 & 0.0 & 0.0 & 5.9 & 18\\ 
  5 & 0.0 & 0.0 & 2.1 & 4.6 & \textbf{87.1} & 0.0 & 0.0 & 0.0 & 6.2  & 32\\ 
  6 & 0.0 & 0.0 & 0.0 & 0.0 & 0.0 & \textbf{100.0} & 0.0 & 0.0 & 0.0 & 1\\ 
  7 & 0.0 & 0.0 & 6.7 & 0.0 & 0.0 & 0.0 & \textbf{84.9} & 2.1 & 6.3  & 48\\ 
  8 & 0.0 & 0.0 & 0.0 & 0.0 & 0.0 & 0.0 & 12.3 & \textbf{83.9} & 3.8 & 35\\ 
  9 & 1.7 & 0.0 & 0.0 & 0.0 & 0.0 & 0.0 & 23.7 & 1.5 & \textbf{73.1} & 39\\ 
   \hline
\end{tabular}
\end{table}

The assessment of the quality of hand-coded data is of paramount importance in the construction of a~classifier, as the input codes (labels) must be accurate. In our approach, we utilise the codes in their original form; however, it is essential to consider the potential for uncertainty in the preparation of training datasets when interpreting the final results. Further research and the development of algorithms that take this into account will be required to address this issue in greater depth.

\subsection{General description of the data}\label{sec-general-desc-data}

In this section, we present information on the quality of input data. The combined dataset contained 200,875 cases out of which 9,200 cases refer to the official dictionary\footnote{Note that this number is larger than the number of codes because for certain codes more than one case is provided.}. \cref{tab-coverage-size} presents the coverage of occupation 6-digits occupation codes by main groups and source. The last row presents information about the number of rows for a~given dataset. Datasets varied in terms of coverage. The most complete was ePraca with coverage close to 80\% of all 6-digit codes within given occupation group. Due to sample size, hand-coded data covered only a~small fraction of available codes. The combined dataset was missing 186 occupation codes\footnote{Detailed information: 1 group: 9, 2 group: 84, 3 group: 46, 4 group: 1, 5 group: 5, 6 group: 1, 7 group: 27, 8 group: 8, 9 group: 5). List of these codes is provided here \url{https://github.com/OJALAB/job-ads-datasets/blob/main/data/codes-not-coveted.csv}.}. It should be noted that for the training dataset the complete dictionary with all 2,911 codes was used. 

\begin{table}[ht!]
\centering
\caption{Coverage of 6 digit codes by main groups and sources (excluding official dictionary)}
\label{tab-coverage-size}
\begin{tabular}{lrrrrrrrr|r}
  \hline
Group & Info+ & GUS & ESCO & KPRM & Hand (1) & Hand (2) & ePraca & All & Cases \\ 
  \hline
  0 & 0.0 & 100.0 & 33.3 & 0.0 & 0.0 & 0.0 & 100.0 & 100.0 & 28 \\ 
  1 & 6.4 & 85.6 & 19.8 & 1.0 & 42.1 & 15.3 & 79.7 & 95.5 & 5,369 \\ 
  2 & 23.8 & 26.0 & 20.7 & 0.6 & 39.3 & 14.4 & 77.1 & 89.4 & 33,441 \\ 
  3 & 35.7 & 48.0 & 20.0 & 0.7 & 29.7 & 4.8 & 83.6 & 92.5 & 22,229 \\ 
  4 & 53.9 & 21.3 & 20.2 & 1.1 & 53.9 & 9.0 & 95.5 & 98.9 & 18,710 \\ 
  5 & 61.4 & 28.9 & 27.1 & 0.0 & 38.0 & 6.6 & 89.2 & 97.0 & 29,167 \\ 
  6 & 69.8 & 57.1 & 12.7 & 0.0 & 11.1 & 0.0 & 85.7 & 98.4 & 1,769 \\ 
  7 & 46.2 & 46.6 & 16.8 & 0.0 & 39.3 & 4.8 & 89.5 & 94.3 & 39,820 \\ 
  8 & 40.8 & 82.4 & 15.2 & 0.0 & 25.3 & 2.1 & 91.0 & 97.9 & 21,761 \\ 
  9 & 4.0 & 15.9 & 16.7 & 0.0 & 47.6 & 1.6 & 96.0 & 96.0 & 28,581 \\ 
   \hline
   N & 7,007 & 1,342 & 2,114 & 2,941 & 9,992 & 1,035 & 167,244 & & 200,875\\ 
   \hline
\end{tabular}
\begin{flushleft}
Source: Own elaboration. Group (1 digit) as in \cref{tab-structure}.
\end{flushleft}
\end{table}

\begin{table}[ht!]
\centering
\caption{Number of words for a~given group for the combined dataset (official dictionary included)}
\label{tab-number-words}
\begin{tabular}{lrrrrrr|r}
  \hline
Group & Min. & 1st Qu. & Median & Mean & 3rd Qu. & Max. & N \\ 
  \hline
  0 & 4.0 & 13.2 & 36.5 & 52.8 & 54.0 & 208.0 & 28 \\ 
  1 & 1.0 & 35.0 & 83.0 & 131.7 & 183.0 & 2,561.0 & 5,369 \\ 
  2 & 1.0 & 30.0 & 59.0 & 107.0 & 141.0 & 2,836.0 & 33,441 \\ 
  3 & 1.0 & 29.0 & 56.0 & 96.6 & 118.0 & 2,792.0 & 22,229 \\ 
  4 & 1.0 & 25.0 & 45.0 & 80.4 & 83.0 & 3,385.0 & 18,710 \\ 
  5 & 1.0 & 20.0 & 32.0 & 50.2 & 55.0 & 3,215.0 & 29,167 \\ 
  6 & 1.0 & 18.0 & 34.0 & 75.7 & 67.0 & 1,229.0 & 1,769 \\ 
  7 & 1.0 & 19.0 & 32.0 & 52.6 & 57.0 & 1,924.0 & 39,820 \\ 
  8 & 1.0 & 22.0 & 36.0 & 56.2 & 62.0 & 1,459.0 & 21,761 \\ 
  9 & 1.0 & 19.0 & 31.0 & 42.3 & 52.0 & 1,686.0 & 28,581 \\ 
   \hline
   Overall & 1 & 22 & 39 & 70 & 75 & 3,385 & 200,875\\
   \hline
\end{tabular}
\begin{flushleft}
Source: Own elaboration. Groups as in \cref{tab-structure}.
\end{flushleft}
\end{table}

Another factor that should be taken into account is the varying number of cases for each code. For instance, the largest number of cases was present for Industrial and craft workers (7) and Professionals (2) with close to 40k and 34k cases respectively. The less represented were Public officials, senior officials and managers (1), Farmers, horticulturists, foresters and fishermen (6), and Armed Forces (0).

Sources varied in terms of a~number of words describing a~given code. \cref{tab-number-words} presents information on the distribution of the number of words in the combined dataset for each group separately. The median number of words was 39 and varied between from 31 for the 9th to 83 for the 1st group. A comparison of the number of words in the descriptions from the two sources reveals that the ePraca descriptions contain, on average, approximately 50 words, while the descriptions from online sources contain more than 120 words on average. The detailed information for each input dataset is presented in the Supplementary Material in Table \cref{tab-words-by-source}. 

Finally, for building the classifier, we have created training and test data according to the following algorithm:

\begin{enumerate}
    \item[Step 1:] Combine datasets: INFO+, KPRM, HAND (10k), HAND (1k), ESCO and ePraca into one dataset.
    \item[Step 2:] Calculate the size of 70\% of cases stratified by source and 6-digit code.
    \item[Step 3:] Strata with only a~single case were removed and used for testing.
    \item[Step 4:] Strata with $>1$ cases were sampled using a~stratified design with sample sizes defined as in Step 2; 70\% for training and 30\% for testing.
    \item[Step 5:] The final training dataset was created by combining 70\% of the dataset from Step 4  with an official dictionary and GUS thesaurus.
    \item[Step 6:] The final test dataset was created by combining the dataset from Step 3 with 30\% sample from Step 4.
\end{enumerate}

\cref{tab-test-train-cases} presents information on the number of cases and 6-digit codes in train and test datasets by source. The training dataset contains close to 143k cases while testing 58k. The main contribution is from the ePraca dataset. The number of codes in the training dataset is equal to the number of codes in the official classification, while in the test data not all codes are available. 

\begin{table}[ht!]
\centering
\caption{Number of cases and 6-digit codes in train and test data by source}
\label{tab-test-train-cases}
\begin{tabular}{lrrrrrr}
  \hline
  & \multicolumn{3}{c}{Train} & \multicolumn{3}{c}{Test} \\
  \hline
 Source & Cases PL & Cases ML & \# Codes & Cases PL & Cases ML & \# Codes \\ 
  \hline
  Official & 9,200 & 22,143 & 2,911 & - & - & - \\ 
  Thesaurus & 1,342 & 2,687 & 1,338 & - & - & - \\ 
  ePraca & 116,879 & 280,461 & 2,213 & 50,365 & 181,648 & 2,468 \\ 
  ESCO & 1,531 & 3,642 & 557 & 583 & 2,318 & 557 \\ 
  Hand 10k & 6,720 & 16,081 & 708 & 3,272 & 12,050 & 1,039 \\ 
  Hand 1k & 632 & 1,504 & 115 & 403 & 1,523 & 226 \\ 
  Info & 5,004 & 12,010 & 996 & 2,003 & 7,012 & 996 \\ 
  KPRM & 2,058 & 4,926 & 12 & 883 & 3,170 & 12 \\
  \hline
  All & 143,366 & 343,454 & 2,911 & 57,509 & 207,721 & 2,625 \\ 
  \hline
\end{tabular}
\footnotesize{\\ Note: Cases PL refers to Polish dataset, Cases ML refers to multilingual (24 languages) dataset}
\end{table}

In order to verify the approach we translated train and test data from English (based on Google Translate) to 22 languages (all UE languages) with Argos Translate. This dataset over 3,4 mln records for training and close to 1,4 mln for test data. In order to avoid overfitting of the multilingual classifier we selected a~stratified random sample of both datasets independently with 10\% and 15\% from each stratum created by 6-digit occupation code and source type. Information about the number of cases for each dataset (train and test are presented in \cref{tab-test-train-cases} while detailed information about the distribution of languages are presented in \cref{tab-multilingual-dataset}.

\begin{table}[ht]
\caption{Distribution of languages in train and test data (in thousands of records) in the multilingual dataset and information about the number of tokens (in milions) on which XLM-Roberta was trained}
\label{tab-multilingual-dataset}
\centering
\resizebox{\linewidth}{!}{
\begin{tabular}{lrrrlrrr}
\hline
\multicolumn{4}{c}{Languages A-I} & \multicolumn{4}{c}{Languages I-U} \\
\hline
Language & Train & Test & XLM-RoBERTa & Language & Train & Test & XLM-RoBERTa \\
\hline
Bulgarian (bg) & 12.3 (3.9\%) & 7.7 (4.0\%) & 5.5 (4.4\%) & Italian (it) & 12.2 (3.9\%) & 7.6 (4.0\%) & 5.0 (4.0\%) \\
Czech (cs) & 13.3 (4.2\%) & 11.1 (5.8\%) & 2.5 (2.0\%) & Lithuanian (lt) & 13.8 (4.3\%) & 7.2 (3.8\%) & 1.8 (1.5\%) \\
Danish (da) & 15.4 (4.9\%) & 7.4 (3.9\%) & 7.8 (6.2\%) & Latvian (lv) & 14.4 (4.6\%) & 8.1 (4.2\%) & 1.2 (1.0\%) \\
German (de) & 15.7 (4.9\%) & 8.0 (4.2\%) & 10.3 (8.2\%) & Dutch (nl) & 12.9 (4.1\%) & 7.6 (4.0\%) & 5.0 (4.0\%) \\
Greek (el) & 14.5 (4.6\%) & 8.3 (4.3\%) & 4.3 (3.4\%) & Polish (pl) & 12.2 (3.9\%) & 7.3 (3.8\%) & 6.5 (5.2\%) \\
English (en) & 14.2 (4.5\%) & 9.3 (4.8\%) & 55.6 (44.3\%) & Portuguese (pt) & 14.4 (4.5\%) & 10.2 (5.3\%) & 8.4 (6.7\%) \\
Spanish (es) & 16.0 (5.1\%) & 7.9 (4.1\%) & 9.4 (7.5\%) & Romanian (ro) & 12.3 (3.9\%) & 9.8 (5.1\%) & 10.4 (8.3\%) \\
Estonian (et) & 16.5 (5.2\%) & 7.7 (4.0\%) & 0.8 (0.7\%) & Russian (ru) & 12.5 (4.0\%) & 9.5 (5.0\%) & 23.4 (18.7\%) \\
Finnish (fi) & 14.6 (4.6\%) & 7.9 (4.1\%) & 6.7 (5.4\%) & Slovak (sk) & 16.4 (5.2\%) & 10.5 (5.5\%) & 3.5 (2.8\%) \\
French (fr) & 18.8 (5.9\%) & 11.3 (5.9\%) & 9.8 (7.8\%) & Slovenian (sl) & 13.4 (4.2\%) & 8.2 (4.3\%) & 1.7 (1.3\%) \\
Irish (ga) & 14.2 (4.5\%) & 8.0 (4.2\%) & 0.1 (0.1\%) & Swedish (sv) & 14.1 (4.4\%) & 10.2 (5.4\%) & 0.1 (0.1\%) \\
Hungarian (hu) & 15.0 (4.7\%) & 9.3 (4.9\%) & 7.8 (6.2\%) & Ukrainian (uk) & 14.4 (4.5\%) & 7.8 (4.1\%) & 0.0 (0.0\%) \\
\hline
\end{tabular}
}
\end{table}

The number of cases for a~given language varies between the two sources. To illustrate, in the case of hand-coded test data, Polish was the least represented language, with 287 cases, while French was the most represented, with 1,348.  

\cref{tab-multilingual-dataset} contains information about the number of tokens used to train XLM-RoBERTa. The primary objective is to demonstrate that some languages are markedly under-represented in the XLM-RoBERTa dataset, including Estonian, Irish, Swedish and Ukrainian, while others are over-represented, such as German and Russian. This may suggest that the quality of classification may vary between languages.

\section{Experimental setup}\label{sec-setup}

\subsection{Models and implementation}

In the empirical study, we considered the following model variants of bottom-up and top-down approaches described in~\cref{sec:modeling-class-hierarchy}:

\begin{itemize}
    \item As a~baseline, we use bottom-up and top-down with simple linear models (i.e. logistic regression with linear part) for modelling probabilities using TF-IDF representations of texts. 
    \item We compare it with the transformer-based architecture as described in \cref{sec:proposed-architecture}, in both bottom-up and top-down variants fine-tuning them based on the following pre-trained models:
    \begin{itemize}
        \item HerBERT-base (\url{https://huggingface.co/allegro/herbert-base-cased}, with 110M parameters) and HerBERT-large (\url{https://huggingface.co/allegro/herbert-large-cased}, with 336M parameters) -- fine-tuned BERT for Polish language developed by Allegro,
        \item XLM-RoBERTa-base (\url{https://huggingface.co/FacebookAI/XLM-roberta-base}, with 279M parameters) and XLM-RoBERTa-large (\url{https://huggingface.co/FacebookAI/XLM-roberta-large}, with 561M paramters) -- multilingual models developed by Facebook/Meta.
    \end{itemize}
\end{itemize}

As the baseline implementation, we use \texttt{napkinXC}~\citep{Jasinska-Kobus_et_al_2020_PLT}, which is an extremely simple and fast library for extreme multi-class and multi-label classification that implements, among the others, linear one-versus-rest multi-class classifier (used for bottom-up model) and hierarchical softmax (for top-down). In order to facilitate the implementation of our approach, we have developed an open-source software tool that enables users to either utilise existing models or train new ones.

For the implementation of the Transformer architecture, we use PyTorch~\citep{pytorch}, PyTorch Lightning~\citep{pytorch_lightning}, and Huggingface Transformers~\citep{wolf2019huggingface} libraries. We fine-tune transformer-based models using AdamW optimizer ~\citep{loshchilov2017decoupled}, 
with the base learning rate equal to $1\mathrm{e-}5$, weight decay to $0.01$ (for all layers except bias terms and layer norms layers), and Adam hyperparameters to $\beta_1 = 0.9, \beta_2 = 0.999, \epsilon = 1\mathrm{e-}8$. We use a~linear learning rate schedule with a~warm-up phase of 50 training steps. The training was done on a~machine with two Nvidia RTX5000 GPUs with 24 GB RAM each. For base models, we used 10 epochs, while for large, only 5. The reason for that was the computational time, for instance, for \texttt{HerBERT-base} one epoch was about 30 min, for \texttt{XLM-RoBERTa-base} about an hour, for \texttt{HerBERT-large} about 1.5 hours and 
 \texttt{XLM-RoBERTa-large} about 2.5 hours. Training for the top-down approach was approximately two times longer than for the bottom-up approach.

\subsection{Evaluation metrics}

To evaluate the quality of obtained classifiers, we primarily report a~per hierarchy level logistic loss (cross-entropy) to investigate if the proposed approach improves the quality of obtained estimates of conditional probabilities of classes $\eta_c(\bx) \coloneqq \proba(y_c = 1 \given \bx)$ at each level of the hierarchy.
Additionally, we report recall$@k$ at each level of the hierarchy. which indicates how many of all relevant classes were predicted in the set of $k$ predictions coming from a~classifier. Recall is formally defined as:
\begin{equation}
    \operatorname{recall} @ k(\boldsymbol{y}, \boldsymbol{h}(x))=\frac{1}{||\boldsymbol{y}||_1} \sum_{j \in \hat{\mathcal{Y}}_k} \llbracket y_j=1 \rrbracket
\end{equation}
\noindent where $||\boldsymbol{y}||_1$ is $L_1$ norm of vector $\boldsymbol{y}$ -- number of positive classes, $\hat{\mathcal{Y}}_k$ is a~set of $k$ labels with top conditional probabilities $\eta_c(\bx)$ predicted by $ \boldsymbol{h}$ for $\boldsymbol{x}$.
We calculate recall$@k$ separately for each level. Notice that in this case, $||\boldsymbol{y}||_1$ is always equal to 1, as only one class is relevant per hierarchy level. Because of that, in this case, recall$@1$ is indifferent to accuracy. It is also easy to see that the optimal decision rule when predicting for recall$@k$ is to select top-$k$ classes with the highest conditional probabilities $\eta_c(\bx)$~\citep{Lapin_et_al_2018}. This makes recall$@k$ align with our goal of measuring the quality of probability estimates while being more intuitive than logistic loss.

\section{Results}\label{sec-results-classification}

\subsection{Results for country specific classifier (trained on Polish language only)} \label{sec-results-overall}

\begin{figure}
    \centering
    \includegraphics[width=\textwidth]{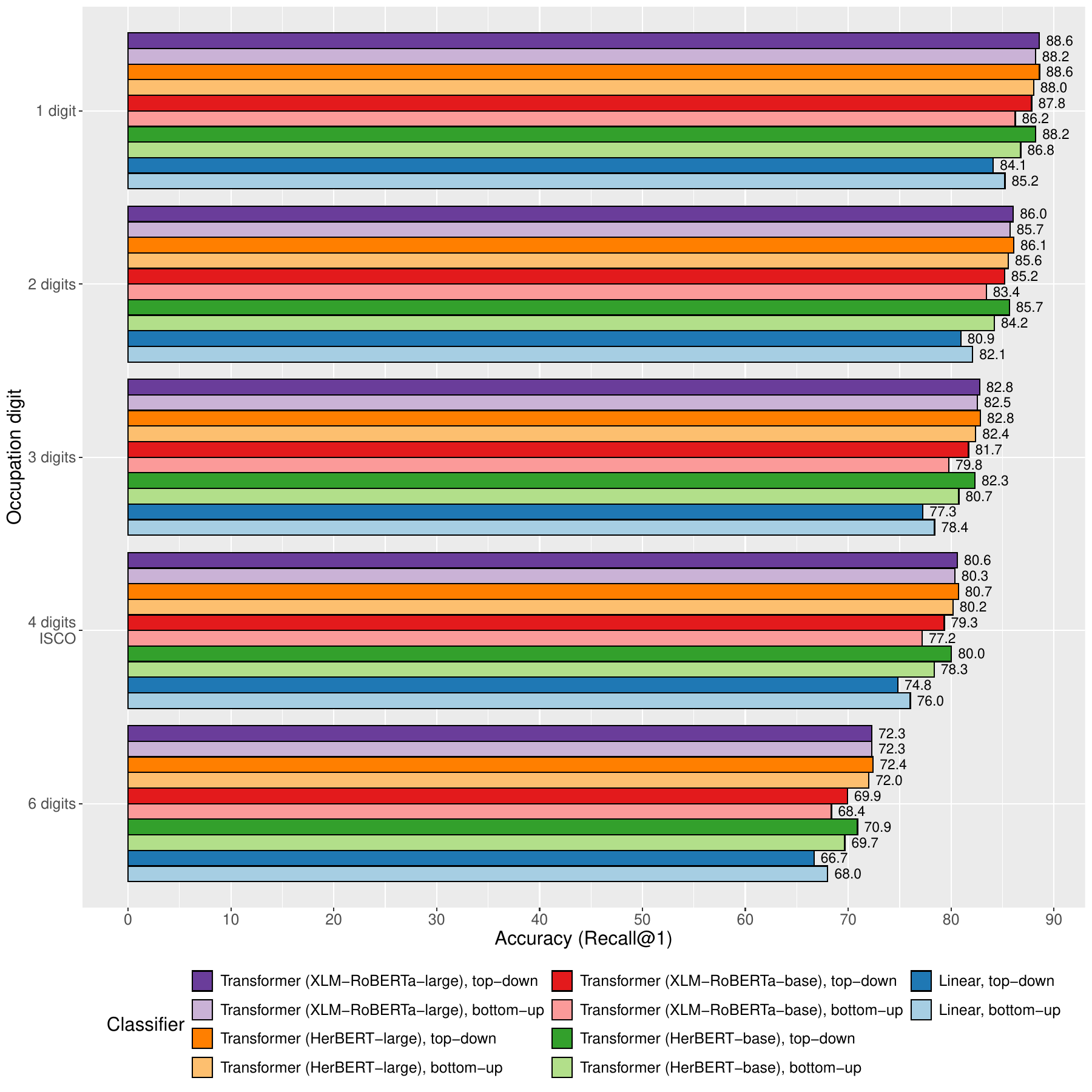}
    \caption{Results for the recall@1 classification of Polish job advertisements by classifier (Overall)}
    \label{fig-recall1a}
\end{figure}

\begin{figure}
    \centering
    \includegraphics[width=\textwidth]{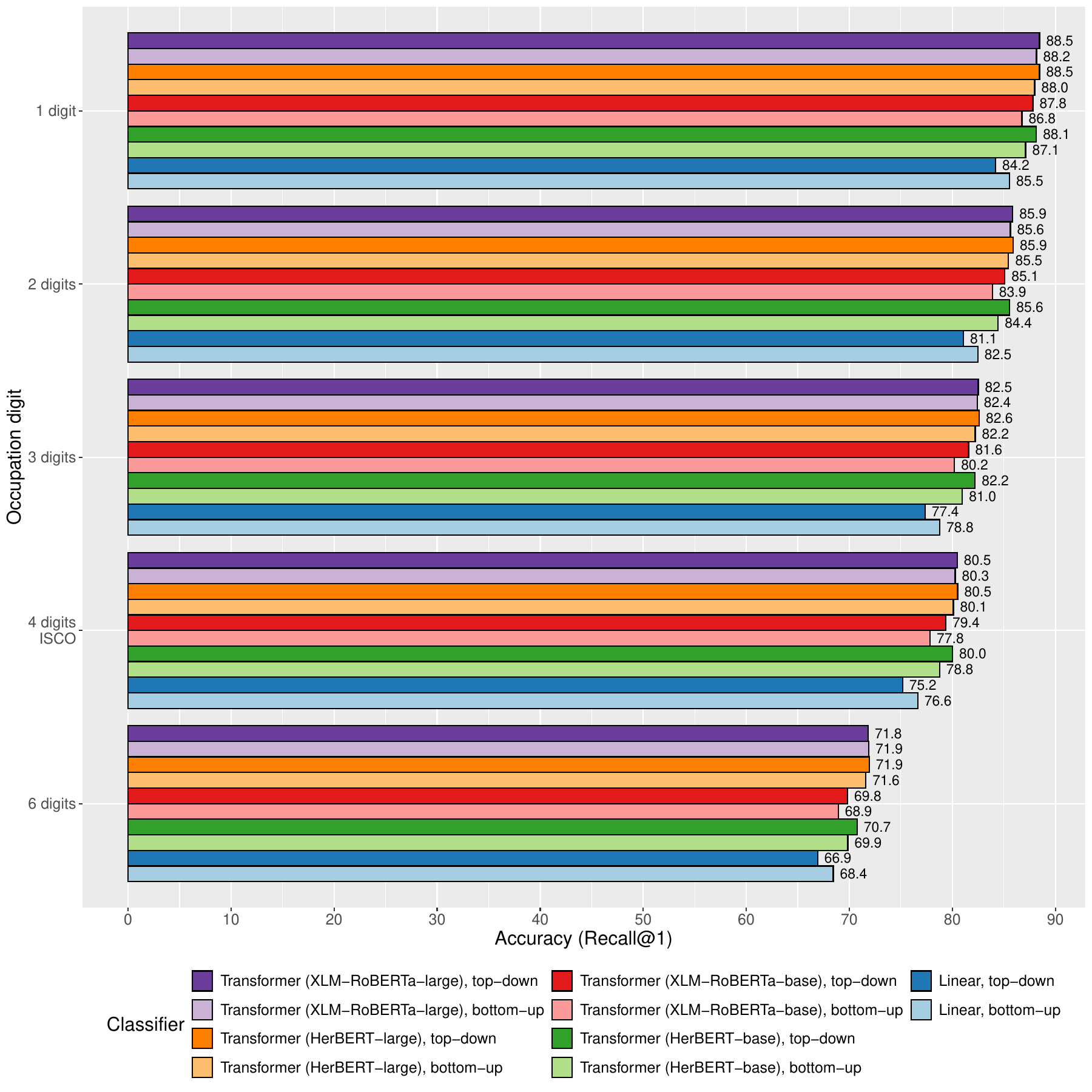}
    \caption{Results for the recall@1 classification of Polish job advertisements by classifier (ePraca only)}
    \label{fig-recall1b}
\end{figure}

\begin{figure}
    \centering
    \includegraphics[width=\textwidth]{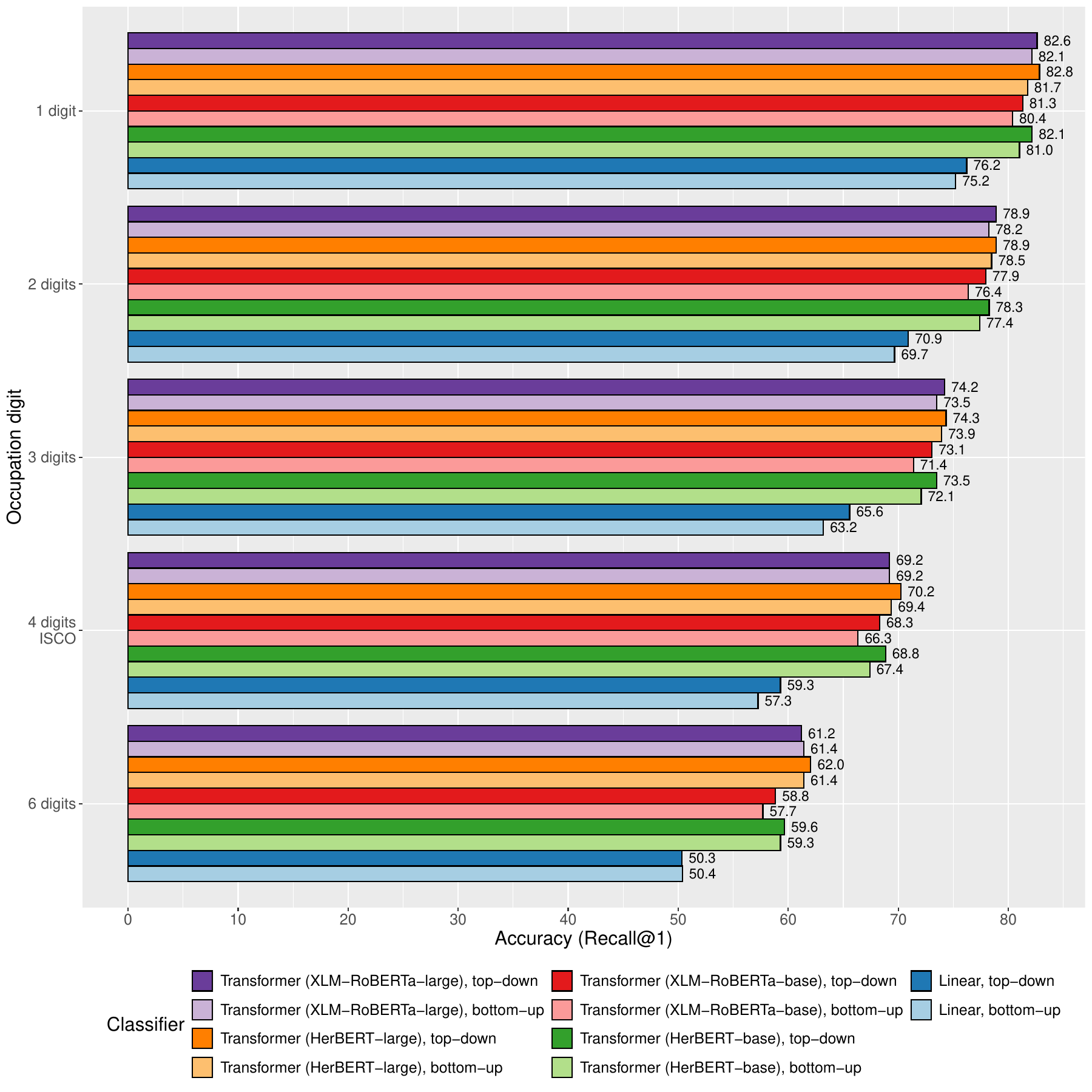}
    \caption{Results for the recall@1 classification of Polish job advertisements by classifier (Hand-coded only)}
    \label{fig-recall1c}
\end{figure}

Figures \ref{fig-recall1a}-\ref{fig-recall1c} compares the accuracy of different classification approaches for job codes across various levels of the KZIS occupation classification system, ranging from 1-digit to 6-digit codes, with ISCO classification at the 4-digit level. Across all datasets (Overall, ePraca, and Hand-coded job offers), transformer-based models consistently outperform linear models. Larger transformer models (HerBERT-large and XLM-RoBERTa-large) achieve the highest accuracy scores across all levels and datasets but are slightly better than base models. 

Notably, the top-down approach consistently yields better results than the bottom-up approach for all models and coding levels. This improvement is particularly pronounced for hand-coded data, indicating that taking the hierarchical structure of occupational codes into account during model training significantly enhances prediction accuracy. For instance, for the hand-coded dataset for HerBERT-base the difference at ISCO level is close to 1.5p.p. The same relation is present for the linear model where the differences between top-down and bottom-up are close to 2 p.p. (e.g., ISCO level)

As expected, the accuracy of all models decreases as the level of detail in the occupational codes increases. For example, the best-performing model (XLM-RoBERTa-large and HerBERT-large with top-down approach) achieves around 88.6\% accuracy for 1-digit codes but drops to about 61\% for 6-digit codes. This trend is consistent across all models and approaches, reflecting the increasing difficulty of classification as the categories become more specific.

Results for country-specific models suggest that large models perform similarly as base ones and country-specific BERT models performs slightly better than the multilingual BERT model. It should be noted that the XLM-RoBERTa-base model consist of 279M parameters while HerBERT 110M so comparison between these models it not entirely fair. However, it suggest that a~smaller model for a~specific language may perform similarly as larger model trained on multi-language dataset. In the next section, we will compare the multilingual model where the dataset contains two languages -- Polish and English.

\subsection{Results for bilingual classifier (trained on Polish and English languages)}

\subsubsection{Overall results}

\begin{figure}[ht!]
    \centering
    \includegraphics[width=\textwidth]{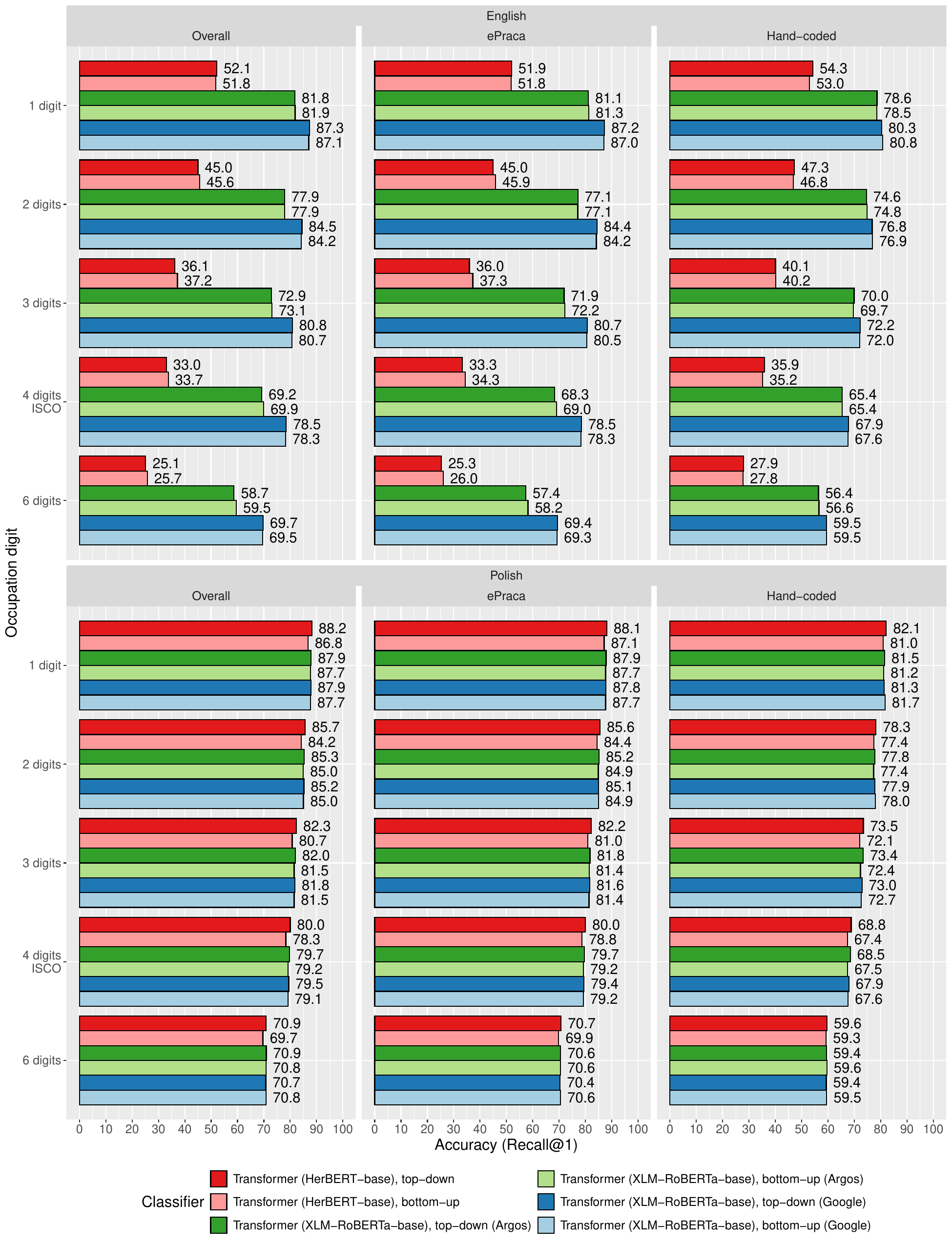}
    \caption{Results for the recall@1 classification of Polish job advertisements by multilingual classifier (XLM-RoBERTa-base)}
    \label{fig-recall2}
\end{figure}

\cref{fig-recall2} compares the accuracy (measured as Recall@1) of different classifiers for occupational coding across various levels of detail, from 1-digit to 6-digit codes, as well as ISCO codes. The classifiers include different transformer models (HerBERT and XLM-RoBERTa) using both top-down and bottom-up approaches. The plot also compares these automated methods to hand-coded results.

The results show that across all levels of occupational coding detail, the models using Google Translate consistently outperform those using Argos Translate. This difference is particularly noticeable for more detailed coding levels. For example, at the 6-digit level, the best-performing Google model (Polish Transformer XLM-RoBERTa-base, bottom-up) achieves about 88\% accuracy, while the corresponding Google Translate model achieves only about 86.4\% accuracy. Notably, the HerBERT model, which was specifically designed for Polish, performs poorly when applied to English text, highlighting the importance of using language-appropriate models.

Interestingly, the Polish versions of the models tend to perform slightly better than their English counterparts for both Argos and Google translations. This suggests that maintaining the original language (Polish in this case) or using a~closely related language for translation might preserve more relevant information for occupational coding tasks. The poor performance of HerBERT on English text further supports this observation, emphasizing the need for language-specific or truly multilingual models in cross-lingual tasks.

As observed in the previous plot, the top-down approach generally yields slightly better results than the bottom-up approach, although the difference is less pronounced here. The automated methods significantly outperform hand-coding across all levels of detail, with the gap widening for more granular classification tasks. The substantial difference in performance between Argos and Google translations highlights the importance of choosing appropriate translation tools in multilingual occupational coding tasks, with the closed-source Google Translate showing superior results in this context. This finding, combined with the language-specific performance differences, underscores the complexity of cross-lingual occupational coding and the need for careful selection of both translation tools and language models.

\subsubsection{Detailed results for the hand-coded data}

This section presents a~more detailed analysis of the combined hand-coded data set, comprising 10,000 and 1,000 records. The data primarily come from online sources, and the classifier will be employed for further analysis of this type of data. \cref{tab-conf-transformer} contains the corresponding data for the transformer model. The results are presented separately for the training and test data. For the training dataset, the transformer model achieves over 92\% agreement for each group, while the logistic model performs significantly worse for the 1, 3, 4, and 9 groups.

\begin{table}[ht!]
\centering
\caption{Confusion matrix for 1 digit Occupation Groups (codes) for the transformer model with top-down approach on the test data}
\label{tab-conf-transformer}
\small
\begin{tabular}{rrrrrrrrrrr}
 \hline
 \hline
  \multicolumn{11}{c}{\textbf{Country specific model -- HerBERT}} \\
  \hline
  \hline
  \multicolumn{11}{c}{Base model (110M parameters)} \\
  \hline
True & 1 & 2 & 3 & 4 & 5 & 6 & 7 & 8 & 9 & N \\ 
  \hline
1 & \textbf{81.0} & 11.6 & 2.6 & 1.7 & 2.2 & 0.0 & 0.9 & 0.0 & 0.0    & 232 \\
2 & 3.1 & \textbf{89.1} & 4.4 & 1.5 & 1.6 & 0.0 & 0.3 & 0.0 & 0.0     & 1,273 \\
3 & 2.6 & 15.7 & \textbf{67.2} & 2.6 & 4.8 & 0.0 & 5.2 & 1.0 & 1.0    & 503 \\
4 & 0.8 & 12.8 & 4.5 & \textbf{70.9} & 3.0 & 0.0 & 0.0 & 1.5 & 6.4    & 265 \\
5 & 1.8 & 4.0 & 4.5 & 1.5 & \textbf{85.7} & 0.0 & 0.8 & 0.0 & 1.8     & 399 \\
6 & 14.3 & 14.3 & 0.0 & 0.0 & 0.0 & \textbf{57.1} & 0.0 & 0.0 & 14.3  & 7 \\
7 & 0.8 & 1.4 & 2.9 & 0.0 & 0.2 & 0.0 & \textbf{89.2} & 2.7 & 2.9     & 518 \\
8 & 0.0 & 0.4 & 0.8 & 1.6 & 0.8 & 0.0 & 17.8 & \textbf{76.4} & 2.3    & 258 \\
9 & 0.0 & 0.9 & 2.7 & 4.5 & 3.2 & 1.4 & 7.7 & 0.9 & \textbf{78.6}     & 220 \\
  \hline
  \multicolumn{11}{c}{Large model (340M parameters)} \\
  \hline
1 & \textbf{82.8} & 11.2 & 1.7 & 1.7 & 2.2 & 0.0 & 0.4 & 0.0 & 0.0  & 232 \\
2 & 3.5 & \textbf{88.2} & 4.8 & 1.6 & 1.6 & 0.0 & 0.3 & 0.0 & 0.0& 1,273 \\
3 & 2.0 & 13.9 & \textbf{68.8} & 2.6 & 4.4 & 0.0 & 6.0 & 1.2 & 1.2  & 503 \\
4 & 1.1 & 10.2 & 6.0 & \textbf{71.3} & 3.0 & 0.0 & 0.0 & 1.1 & 7.2  & 265 \\
5 & 1.5 & 4.3 & 6.3 & 1.0 & \textbf{85.0} & 0.0 & 0.8 & 0.0 & 1.3  & 399 \\
6 & 0.0 & 14.3 & 0.0 & 0.0 & 0.0 & \textbf{71.4} & 0.0 & 0.0 & 14.3  & 7 \\
7 & 0.6 & 1.4 & 1.7 & 0.2 & 0.4 & 0.0 & \textbf{90.7} & 2.7 & 2.3  & 518 \\
8 & 0.0 & 0.0 & 1.9 & 1.2 & 0.4 & 0.0 & 18.2 & \textbf{77.9} & 0.4  & 258 \\
9 & 0.5 & 0.0 & 2.7 & 5.0 & 2.7 & 2.3 & 9.1 & 0.9 & \textbf{76.8}  & 220 \\
   \hline
   \hline
  \multicolumn{11}{c}{\textbf{Multilingual model -- XLM-RoBERTa-base}} \\
  \hline
  \hline
  \multicolumn{11}{c}{Polish language} \\
  \hline
  & 1 & 2 & 3 & 4 & 5 & 6 & 7 & 8 & 9 \\ 
  \hline
1 & \textbf{76.7} & 13.8 & 5.2 & 1.7 & 1.7 & 0.0 & 0.9 & 0.0 & 0.0   & 232 \\
2 & 3.5 & \textbf{87.0} & 5.3 & 2.0 & 1.6 & 0.0 & 0.5 & 0.0 & 0.0    & 1,273 \\
3 & 2.2 & 14.1 & \textbf{67.0} & 3.2 & 6.4 & 0.0 & 4.6 & 1.4 & 1.2   & 503 \\
4 & 1.1 & 10.9 & 6.0 & \textbf{69.8} & 1.9 & 0.0 & 0.0 & 2.6 & 7.5   & 265 \\
5 & 1.5 & 3.5 & 6.5 & 1.3 & \textbf{85.2} & 0.0 & 0.3 & 0.0 & 1.8    & 399 \\
6 & 0.0 & 14.3 & 0.0 & 0.0 & 0.0 & \textbf{57.1} & 0.0 & 0.0 & 28.6  & 7 \\
7 & 0.8 & 1.4 & 3.1 & 0.0 & 0.2 & 0.0 & \textbf{87.3} & 4.1 & 3.3    & 518 \\
8 & 0.0 & 0.8 & 0.4 & 1.6 & 0.4 & 0.0 & 19.0 & \textbf{76.7} & 1.2   & 258 \\
9 & 0.5 & 0.5 & 2.3 & 6.4 & 3.2 & 1.4 & 7.3 & 1.4 & \textbf{77.3}     & 220 \\
  \hline
  \multicolumn{11}{c}{English language} \\
  \hline
1 & \textbf{78.9} & 14.1 & 3.0 & 1.5 & 1.5 & 0.0 & 1.0 & 0.0 & 0.0   & 232 \\
2 & 3.9 & \textbf{85.1} & 6.9 & 2.0 & 1.6 & 0.0 & 0.6 & 0.0 & 0.0 & 1,273 \\
3 & 2.4 & 12.6 & \textbf{67.8} & 3.9 & 5.0 & 0.0 & 5.4 & 0.9 & 2.0   & 503 \\
4 & 1.6 & 10.5 & 4.3 & \textbf{69.9} & 1.6 & 0.0 & 0.8 & 3.9 & 7.4   & 265 \\
5 & 1.3 & 4.2 & 9.1 & 1.8 & \textbf{81.3} & 0.0 & 1.0 & 0.3 & 1.0   & 399 \\
6 & 0.0 & 14.3 & 0.0 & 0.0 & 0.0 & \textbf{57.1} & 0.0 & 0.0 & 28.6   & 7 \\
7 & 0.6 & 1.2 & 3.1 & 0.2 & 0.2 & 0.0 & \textbf{87.6} & 3.9 & 3.3   & 518 \\
8 & 0.0 & 0.4 & 0.8 & 1.6 & 0.4 & 0.0 & 17.3 & \textbf{77.5} & 2.0   & 258 \\
9 & 0.5 & 0.5 & 1.8 & 6.4 & 1.8 & 0.9 & 9.2 & 1.4 & \textbf{77.5}   & 220 \\
   \hline
\end{tabular}
\end{table}

In order to evaluate the efficacy of a~given approach, it is essential to consider the results of the tests conducted. Once again, the transformer model demonstrated superior performance for all groups, with a~precision rate exceeding 68\% (with the exception of Occupation Group 6). In contrast, the logistic model exhibited lower precision rates for some groups, with a~difference of approximately 10 percentage points. For instance, the logistic model for Group 3 achieved a~precision rate of 57.5\%, while the transformer model surpassed this rate, reaching over 68\%. These results indicate that the transformer model is the optimal choice. Furthermore, we emphasize that the model's performance can be validated using out-of-sample data, particularly new advertisements and clerical reviews with expert coding.

The confusion matrices for the XLM-RoBERTa-base transformer model reveal its performance in occupational coding tasks across Polish and English languages. Overall, the model demonstrates high accuracy, with the majority of predictions correctly falling on the diagonal for both languages. Notably, Occupation Groups 2, 5, and 7 consistently show over 80\% correct classifications, indicating the model's strength in identifying these occupational groups. However, some common misclassification patterns emerge, such as Group 2 being mistaken for Group 1, and Group 3 for Group 2, suggesting potential similarities or overlaps in these occupational definitions that challenge the model's discriminative capabilities.

Interestingly, while the overall performance is comparable between Polish and English, some nuanced differences are observed. For instance, Occupation Group 1 is more accurately classified in English (78.9\%) compared to Polish (76.7\%), whereas Group 5 shows superior performance in Polish (85.2\%) over English (81.3\%). These variations highlight the potential impact of language-specific nuances on occupational coding accuracy, emphasizing the importance of considering linguistic context in model development and application.

The matrices also reveal challenges in classifying low-frequency categories, exemplified by Occupation Group 6, which has only 7 samples and exhibits poor performance with a~high misclassification rate to Group 9. This underscores the persistent challenge in machine learning of accurately handling rare or underrepresented classes, particularly in specialized domains like occupational coding.

Despite these challenges, the consistency in misclassification patterns across both languages suggests that many errors stem from inherent similarities between certain occupational categories rather than language-specific issues. This insight is valuable for future refinements of occupational classification systems and the development of more robust coding models. The strong overall performance of the XLM-RoBERTa-base model across languages demonstrates its potential as a~powerful tool for automated occupational coding, while also highlighting areas for future improvement, particularly in distinguishing between similar categories and handling low-frequency occupations. Detailed results are presented in Supplementary Materials.

\subsection{Results for multilingual dataset (24 languages)}

In this section, we present the results for a~multilingual dataset that has been trained with XLM-RoBERTa models. \cref{tab-multi-basic-result} presents the results for the test dataset only. The table also includes two approaches to modelling the hierarchy for 1, 2, 4 (ISCO) and 6 (KZiS) digits. The results are comparable to those observed for the Polish and bilingual models. The overall classification rate for 1-digit codes is approximately 84\% (approximately 2-4 p.p. lower than the first two approaches), and the hand-coded classification rate is approximately 78\%. As previously noted, the inclusion of the hierarchy improves predictions for the hand-coded data as the hierarchy is traversed.

\begin{table}[ht]
\centering
\caption{Recall@1 metric for the multilingual test data by dataset and transformer approach}
\label{tab-multi-basic-result}
\begin{tabular}{llrrrr}
  \hline
Dataset & Hierarchy & 1 digit & 2 digits & ISCO (4 digits) & KZiS (6 digits) \\ 
  \hline
Overall & bottom-up & 84.37 & 80.87 & 73.67 & 64.42 \\ 
 & top-down & 84.40 & 80.70 & 73.25 & 63.74 \\ 
  ePraca & bottom-up & 84.38 & 80.88 & 73.77 & 64.41 \\ 
   & top-down & 84.43 & 80.75 & 73.46 & 63.90 \\ 
  Hand-coded & bottom-up & 77.82 & 72.95 & 62.37 & 53.51 \\ 
   & top-down & 78.59 & 73.62 & 62.46 & 53.15 \\ 
   \hline
\end{tabular}
\end{table}

\begin{figure}[ht!]
    \centering
    \includegraphics[width=\textwidth]{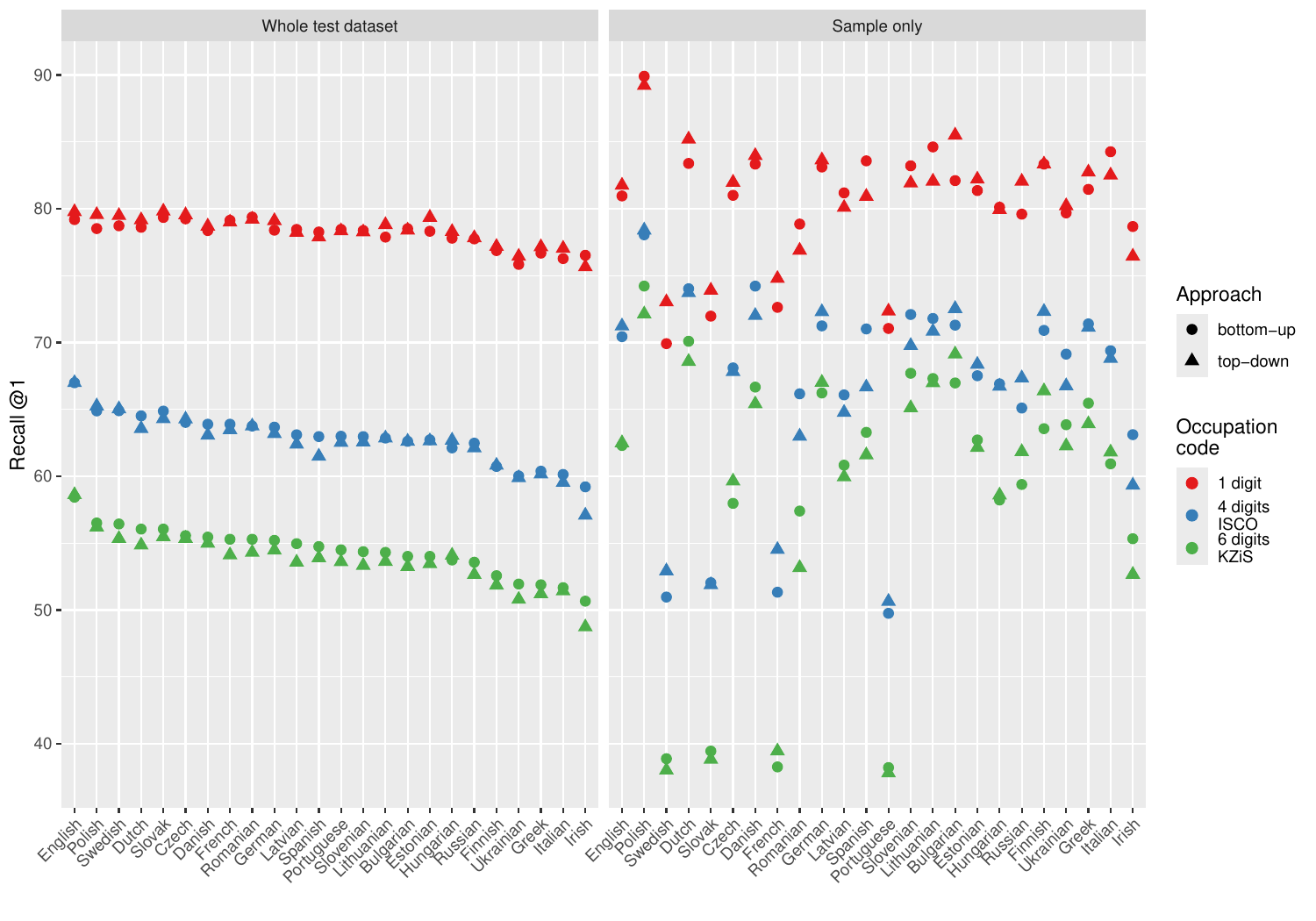}
    \caption{Results for the multilingual model for 1-digit, 4-digits and 6-digit codes by the approach (bottom-up and top-down) for hand-coded test data (whole and sampled). Note X axis is sorted according to the results for the whole data.}
    \label{fig-multi-results}
\end{figure}

As indicated in \cref{sec-general-desc-data}, the sample size for languages differs considerably in the test data. Accordingly, in order to evaluate the accuracy of the multilingual classifier, we present the results for each translated test data set, including the complete data set and the sample utilized for training. \cref{fig-multi-results} depicts the accuracy (recall@1) for each language across 1-digit, 4-digit (ISCO) and 6-digit (KZiS) occupation codes, employing both the bottom-up and top-down approaches. The figures illustrate the results for the complete and sampled hand-coded test data sets. 

As expected, the highest classification was observed for English (translated with Google Translate, the most prevalent language in the XLM-RoBERTa dataset) and Polish (original), with a~near 60\% accuracy at 6-digit codes and a~near 80\% accuracy at 1-digit codes. The top-down method yielded improvements in classification for 67\% of languages at the 1-digit level and 20\% at the ISCO level. The top-down approach yielded an average improvement of 0.3 p.p. at the 1-digit code level, while exhibiting a~decline of -0.3 p.p. at the 4-digit code level. Notably, this approach consistently enhanced classification outcomes for Polish and English languages, suggesting that the impact on other languages may be diminished due to potential discrepancies in translation from English to other languages via the Argos library. Moreover, the degree of accuracy at the 6-digit level exhibits a~notable disparity, with a~near 60\% success rate for English and a~figure approaching 50\% for Irish. This trend persists at the ISCO level, while the variance at the 1-digit level is less pronounced. 

The results for the sampled part of the test data diverge significantly from the results for the entire dataset. For instance, the French result for 6-digit codes was approximately 40\%, while the overall test data yielded a~figure exceeding 55\%. Conversely, the Polish and English results exhibited a~reversed pattern, with values of over 80\% and 90\%, respectively, in comparison to approximately 80\% for both.

A modest correlation was observed between the accuracy of the XLM-Roberta model and the coverage of languages. This is evidenced by the Pearson's and Spearman's correlation coefficients between accuracy and the share of tokens, which were 0.26 and 0.17, respectively. The highest Pearson's and Spearman's correlation coefficients were observed for the share of cases in the test data (0.37 and 0.22, respectively). The detailed results for this classifier are presented in the Supplementary Materials.

\section{Discussion}

In this paper, we propose a~classifier and conditional probability estimator of occupation codes for job vacancy statistics that takes into account the hierarchical structure of the occupation classification. Occupations as proxies of job-related skills seem to be especially important in explaining the technological, demographic, and social transformation in the labour markets. The proposed classifier can be used for automated occupational coding by official statistics or labour market institutions that collect unstructured information about labour demand and formulate policies. To our knowledge this is the first publicly available model for 24 languages along with detailed description on the data and predictions quality.

The paper highlighted the importance of data quality as an input to the training and testing models. We have shown that the hand-coded and administrative data can introduce additional uncertainty due to misclassification of labels and hierarchical classification.  This process should certainly be included in the National Statistical Quality Assurance Frameworks and the European Statistics Code of Practice in order to progressively include indicators on the quality of training and testing data used for official statistics.

Our approach can be used with information-abundant online and administrative data to enrich survey-based job vacancy data. This is especially important nowadays, when survey research faces the problems of non-response and under-reporting. The use of online job advertisements is the main job search method and often also recruitment. They contain plenty of unstructured information. While statistical offices lack methods to classify such text, we provide one. Our approach is dedicated to job offers, and our procedure shows how to incorporate country-specificity on the example of Poland. 

We use the Polish Classification of occupations and specializations, which is six digits, going one level further than the international ISCO classification, and is similar to the ESCO classification. We compare models trained on English and Polish languages and show the limitations of applying generic methods based on English language. We find that the country-specific model performs better than the multilingual one. This confirms the need for language-specific models, as it outperformed the generic ones that were only prepared for our task, and not trained with Polish language. However, there were examples of occupations (e.g., managers) that were well classified in English. This might result from the fact that some occupational titles have English origin and are often used in English form in job advertisements. This may be an example for other language, and institution-specific countries.

In this paper, our focus is on the Polish KZiS and ISCO. However, it should be noted that other classifications may be of interest to users, including O*NET, ESCO and SOC. One potential avenue for investigation is the utilisation of publications that align with one another in terms of their respective classifications. For example, Eurostat has published a~report entitled "The crosswalk between ESCO and O*NET" (see \url{https://esco.ec.europa.eu/en/about-esco/data-science-and-esco/crosswalk-between-esco-and-onet}), which may be used to align our predictions for ISCO with the O*NET classification system. An alternative approach may be to draw upon the work of \citet{stkechly2023propozycja}, who establishes a~link between the Polish KZiS and ESCO classifications. Ultimately, researchers may utilise the pre-trained models presented here and develop their own classifiers.

\section*{Acknowledgements}

We would like to thank Natalie Shlomo and Piotr Chlebicki for their valuable feedback on the previous draft of this paper. 

Cherniaiev Herman, Pater Robert, Wydmuch Marek work and research were conducted in the Educational Research Institute -- National Research Institute (IBE-PIB) in Warsaw, and funded by the Polish Ministry of Education and Science within a~project \textit{Supporting IQS functioning and improvements in order to use its solutions in achieving the country's development strategy aims}, ZSK5 (POWR.02.13.00-IP.02-00-002/20). The preparation of hand-coded datasets was financed by IBE. Maciej Be\-rę\-se\-wicz was partially supported by the Polish National Agency for Academic Exchange (NAWA) under the project number BPN/BEK/2023/1/00099. In addition Cherniaiev Herman, Pater Robert and Maciej Beręsewicz work supported by National Science Centre, Poland within a project OJALAB: Online job advertisements to study skill demand and job search patterns [Grant number: 2024/53/B/HS4/01580].

Marek Wydmuch and Maciej Beręsewicz contributed equally to the paper. Marek Wydmuch was responsible for developing and coding the classifier, Maciej Beręsewicz was responsible for preparing training and testing datasets, computation, preparation of the manuscript, and assessing the results. Herman Cherniaiev collected and initially processed the data on job offers. Robert Pater reviewed the literature and provided the economic and the educational context of research.

Information on software and models 

\begin{itemize}
\small
    \item Codes for the paper and the project repository
    \begin{itemize}
        \item \url{https://github.com/OJALAB/paper-job-ads-classifier}
        \item \url{https://github.com/OJALAB}
    \end{itemize}
    \item Software (source codes for the classifier): 
    \begin{itemize}
        \item \url{https://github.com/OJALAB/job-ads-classifier}
    \end{itemize}
    \item Models (weights): 
    \begin{itemize}
        \item \url{https://repod.icm.edu.pl/dataset.xhtml?persistentId=doi:10.18150/OCUTSI}
    \end{itemize}
    \item Tutorial (installation and prediction): 
    \begin{itemize}
        \item \url{https://colab.research.google.com/drive/1a425aagT0lczRxXPWoUlf5aFxUII37nh?usp=sharing}
    \end{itemize}
\end{itemize}

\clearpage
\theendnotes

\clearpage
\bibliography{bibliography}
\bibliographystyle{apalike}

\clearpage

\appendix

\begin{center}
\Large
    Supplementary materials for the paper \\ 
    \textit{Multilingual hierarchical classification of job advertisements for job vacancy statistics}
\end{center}

\textbf{List of Supplementary Materials}

\begin{enumerate}
    \item Additional information on the data and its quality (Section A).
    \item Models results in Excel Spreadsheet: \texttt{models-accuracy.xlsx}.
    \item A short tutorial on using the software and models for classification and estimation of conditional probability.
\end{enumerate}

\clearpage

\section{Additional information on the data and its quality}

\begin{table}[ht!]
\caption{Number of ads web-scraped between 2021-06-28 and 2021-11-08}
\label{tab-app-sampling-frame}
\centering
\begin{tabular}{rlr}
  \hline
 & Source & N \\ 
  \hline
1 & www.pracuj.pl & 447,624 \\ 
  2 & www.praca.egospodarka.pl & 330,465 \\ 
  3 & www.praca.pl & 303,771 \\ 
  4 & www.aplikuj.pl & 228,575 \\ 
  5 & www.nuzle.pl & 111,623 \\ 
  6 & jobdesk.pl & 106,119 \\ 
  7 & gratka.pl & 91,935 \\ 
  8 & ePraca & 61,596 \\ 
  9 & www.infopraca.pl & 45,164 \\ 
  10 & www.gowork.pl & 24,962 \\ 
  11 & www.jobs.pl & 22,741 \\ 
  12 & www.karierawfinansach.pl & 12,774 \\ 
  13 & www.jober.pl & 9,139 \\ 
  14 & www.absolvent.pl & 2,744 \\ 
  15 & nofluffjobs.com & 2,172 \\ 
  16 & www.goldenline.pl & 2,130 \\ 
  17 & praca.dlastudenta.pl & 2,019 \\ 
  18 & students.pl & 414 \\ 
   \hline
\end{tabular}
\end{table}

\clearpage
\begin{table}[ht!]
\caption{Information on stratification variables and population size (in thousands)}
\label{tab-population-stratas}
\centering
\resizebox{\linewidth}{!}{
\begin{tabular}{rrrrrrrrrr}
  \hline
 & [1,10] & (10,25] & (25,50] & (50,100] & (100,150] & (150,200] & (200,300] & (300,Inf] & $\sum$ \\ 
  \hline
  biurowy & 1.15 & 2.04 & 1.98 & 6.28 & 3.46 & 1.50 & 1.83 & 0.90 & 19.15 \\ 
  (\textit{office}) &   &   &   &   &   &   &   &   &   \\ 
  
  doradcy & 0.24 & 0.34 & 0.50 & 4.75 & 7.46 & 10.25 & 9.68 & 2.07 & 35.30 \\ 
  (\textit{advisors}) &   &   &   &   &   &   &   &   &   \\ 
  
  magazynier & 0.94 & 1.88 & 2.43 & 8.59 & 7.69 & 4.83 & 4.99 & 1.90 & 33.26 \\ 
  (\textit{warehouseman}) &   &   &   &   &   &   &   &   &   \\ 
  
  kucharze & 1.27 & 2.73 & 2.83 & 6.97 & 3.10 & 0.75 & 0.75 & 0.23 & 18.63 \\ 
  (\textit{cooks}) &   &   &   &   &   &   &   &   &   \\ 
  
  produkcja & 0.60 & 1.49 & 1.55 & 5.10 & 5.29 & 4.21 & 4.42 & 2.02 & 24.67 \\ 
  (\textit{production}) &   &   &   &   &   &   &   &   &   \\ 
  
  przedstawiciele & 0.26 & 0.43 & 0.51 & 4.97 & 10.74 & 7.33 & 6.76 & 1.84 & 32.84 \\ 
  (\textit{representatives}) &   &   &   &   &   &   &   &   &   \\ 
  
  specjaliści & 1.27 & 2.15 & 2.62 & 11.53 & 25.73 & 27.40 & 32.59 & 26.82 & 130.11 \\ 
  (\textit{specialists}) &   &   &   &   &   &   &   &   &   \\ 
  
  sprzątacze & 0.56 & 1.40 & 1.24 & 2.60 & 0.70 & 0.54 & 0.00 & 0.00 & 7.04 \\ 
  (\textit{cleaners}) &   &   &   &   &   &   &   &   &   \\ 
  
  sprzedawcy & 1.97 & 4.20 & 5.59 & 16.78 & 11.03 & 5.52 & 7.79 & 1.85 & 54.73 \\ 
  (\textit{sellers}) &   &   &   &   &   &   &   &   &   \\ 
  
  inne & 31.74 & 53.37 & 60.60 & 186.46 & 200.82 & 169.66 & 208.79 & 123.80 & 1,035.24 \\ 
  (\textit{other}) &   &   &   &   &   &   &   &   &   \\ 
  
  $\sum$ & 40.01 & 70.04 & 79.86 & 254.03 & 276.01 & 232.00 & 277.60 & 161.43 & 1,390.98 \\
   \hline
\end{tabular}
}
\end{table}

\clearpage
\begin{table}[ht]
\caption{Sample size by source}
\label{tab-sample-size-source}
\centering
\begin{tabular}{lrr}
  \hline
Source & N & \% \\ 
  \hline
pracuj.pl & 2,201 & 22.0 \\ 
  praca.egospodarka.pl & 1,967 & 19.7 \\ 
  praca.pl & 1,881 & 18.8 \\ 
  aplikuj.pl & 880 & 8.8 \\ 
  nuzle.pl & 764 & 7.6 \\ 
  jobdesk.pl & 757 & 7.6 \\ 
  gratka.pl & 561 & 5.6 \\ 
  ePraca & 241 & 2.4 \\ 
  inne & 750 & 7.5 \\ 
   \hline
   Total & 10,002 & 100\\
   \hline
\end{tabular}
\end{table}

\clearpage
\begin{table}[ht!]
    \centering
    \caption{Phrases used, sample and population number of ads}
    \label{tab-1000-sample}
    \begin{tabular}{lrr}
    \hline
     Phrase    &  Sample ($n$) & Population ($N$)\\
     \hline
     blockchain|block chain    & 10 & 39 \\ 
     (\textit{blockchain|block chain})    &   &   \\ 
     
     cloud|cloud computing|chmura    & 100 &  1,007\\ 
     (\textit{cloud|cloud computing})    &   &   \\ 
     
     cyberbezpiecz    & 100 & 488 \\ 
     (\textit{cybersecurity})    &   &   \\ 
     
     diagnosta laboratoryjny    & 10 & 46 \\ 
     (\textit{laboratory diagnostician})    &   &   \\ 
     
     etyki biznesu    & 9 & 9 \\ 
     (\textit{business ethics})    &   &   \\ 
     
     farmaceuta    & 10 & 168 \\
     (\textit{pharmacist})    &   &   \\ 
     
     (frontend|front end) developer    & 100 &  287\\ 
     (\textit{(frontend|front end) developer})    &   &   \\ 
     
     informatyk(a|i) przemysłow(a|ej)|industrial computer science    & 30 & 30 \\ 
     (\textit{industrial computer science})    &   &   \\ 
     
     internet rzeczy|IoT|internet of things    & 14 & 14 \\ 
     (\textit{IoT|internet of things})    &   &   \\ 
     
     kryminaln|cyberprzest    & 22 & 22 \\ 
     (\textit{criminal|cybercriminals})    &   &   \\ 
     
     lekarz weterynarii    & 10 &  33 \\ 
     (\textit{veterinarian})    &   &   \\ 
     
     pianino|fortepian|pianina    & 4 & 4 \\ 
     (\textit{piano})    &   &   \\ 
     
     położna    & 10 & 30 \\ 
     (\textit{midwife})    &   &   \\ 
     
     robotyk    & 100 & 1667 \\ 
     (\textit{roboticist|robotics specialist})    &   &   \\ 
     
     spawacz (mag|mit|mig)    & 10 & 219 \\
     (\textit{welder (mag|mit|mig)})    &   &   \\ 
     
     tłumacz przysięgły    & 5 & 5 \\ 
     (\textit{sworn translator})    &   &   \\ 
     
     toksykolog    & 14 & 14 \\ 
     (\textit{toxicologist})    &   &   \\ 
     
     uczenie maszynowe|machine learning|data scientist & 150 & 354\\
     (\textit{machine learning|data scientist})    &   &   \\ 
     
     virtual reality|wirtualna rzeczywostość    & 5 &  7\\ 
     (\textit{virtual reality})    &   &   \\ 
     
     \hline
     unspecified phase (the rest) & 287 & 227,122\\
     \hline
     -- & 1,000 & 227,825\\
     \hline
    \end{tabular}
\end{table}

\clearpage

\begin{table}[ht]
\centering
\caption{Unique 6- and 1-digit codes by strata for hand-coded 1k dataset}
\label{table-hand1k-strata}
\small
\begin{tabular}{lrr}
  \hline
  & \multicolumn{2}{c}{Unique codes} \\
Stratum (Phrase) & 6-digits & 1-digit \\ 
  \hline
other & 138 &   8 \\ 
  robotyk & 30 & 6 \\
  (\textit{roboticist|robotics specialist}) & & \\
  cyberbezpiecz & 29 & 3 \\
  (\textit{cybersecurity}) & & \\
  cloud$|$cloud computing$|$chmura & 24 & 4 \\
  (\textit{cloud|cloud computing}) & & \\
  uczenie maszynowe$|$machine learning$|$data scientist & 21 & 5 \\
  (\textit{machine learning|data scientist}) & & \\
  informatyk(a$|$i) przemysłow(a$|$ej)$|$industrial computer science & 11 & 4 \\
  (\textit{industrial computer science}) & & \\
  blockchain$|$block chain & 9 & 3 \\
  (\textit{blockchain|block chain}) & & \\
  toksykolog & 8 & 3 \\
  (\textit{toxicologist}) & & \\
  etyki biznesu & 8 & 4 \\
  (\textit{business ethics}) & & \\
  lekarz weterynarii & 6 & 2 \\
  (\textit{veterinarian}) & & \\
  internet rzeczy$|$IoT$|$internet of things & 6 & 2 \\
  (\textit{IoT|internet of things}) & & \\
  kryminaln$|$cyberprzest & 6 & 1 \\
  (\textit{criminal|cybercriminals}) & & \\
  diagnosta laboratoryjny & 5 & 2 \\
  (\textit{laboratory diagnostician}) & & \\
  farmaceuta & 5 & 3 \\
  (\textit{pharmacist}) & & \\
  (frontend$|$front end) developer & 4 & 1 \\
  (\textit{(frontend|front end) developer}) & & \\
  położna & 4 & 3 \\
  (\textit{midwife}) & & \\
  pianino$|$fortepian$|$pianina & 4 & 2 \\
  (\textit{piano}) & & \\
  virtual reality$|$wirtualna rzeczywostość & 3 & 2 \\
  (\textit{virtual reality}) & & \\
  spawacz (mag$|$mit$|$mig) & 3 & 1 \\
  (\textit{welder (mag|mit|mig)}) & & \\
  tłumacz przysięgły & 1 & 1 \\
  (\textit{sworn translator}) & & \\
   \hline
\end{tabular}
\begin{flushleft}
\begin{small} 
Note: \cref{table-hand1k-strata} provides information on the number of unique 6-digit and 1-digit codes for a~given stratum defined by phrase. This can provide information about the quality of the initial stratification based on keyword searches. For example, the main codes for 'pianino|fortepian|pianina' were related to kitchen and restaurant organisation. The reason for this is that these keywords are related to the name of the restaurant rather than the pianino itself.
\end{small}
\end{flushleft}
\end{table}

\clearpage
\begin{spacing}{1.0} 
\begin{longtable}{lrrrrrrrr}
\caption{Number of words by source and main groups\label{tab-words-by-source}} \\ 
\hline
Source & Group & Min & Q1 & Med & Mean & Q3 & Max & N \\ 
\hline
\endfirsthead
\multicolumn{9}{c}{(Continued from previous page)} \\
\hline
Source & Group & Min & Q1 & Med & Mean & Q3 & Max & N \\ 
\hline
\endhead
\hline
\multicolumn{9}{r}{(Continued on next page)} \\
\endfoot
\hline
\multicolumn{9}{l}{Min = Minimum, Q1 = 1st Quartile, Med = Median,} \\
\multicolumn{9}{l}{Q3 = 3rd Quartile, Max = Maximum, N = Number of observations} \\
\endlastfoot

ePraca & 0 & 11 & 26 & 44 & 55.2 & 52 & 208 & 9 \\ 
 & 1 & 2 & 38 & 66 & 99.8 & 118 & 2,561 & 3,043 \\ 
 & 2 & 2 & 27 & 46 & 74.9 & 84 & 2,310 & 22,920 \\ 
 & 3 & 1 & 29 & 50 & 77.0 & 89 & 1,406 & 16,152 \\ 
 & 4 & 1 & 24 & 42 & 73.0 & 74 & 3,385 & 17,144 \\ 
 & 5 & 1 & 19 & 31 & 40.3 & 49 & 3,215 & 26,434 \\ 
 & 6 & 2 & 18 & 29.5 & 35.8 & 47 & 283 & 1,184 \\ 
 & 7 & 1 & 19 & 30 & 40.1 & 50 & 1,174 & 34,769 \\ 
 & 8 & 2 & 22 & 34 & 42.3 & 54 & 1,241 & 18,243 \\ 
 & 9 & 1 & 19 & 31 & 40.0 & 50 & 1,067 & 27,346 \\ 
\hline
ESCO & 0 & 4 & 6.2 & 18 & 44.5 & 56.2 & 138 & 4 \\ 
 & 1 & 1 & 4 & 22 & 67.9 & 88.5 & 1,233 & 156 \\ 
 & 2 & 1 & 4 & 30 & 63.8 & 88 & 1,035 & 681 \\ 
 & 3 & 1 & 4 & 32 & 61.2 & 99 & 634 & 425 \\ 
 & 4 & 1 & 3 & 24 & 37.7 & 67.5 & 222 & 67 \\ 
 & 5 & 1 & 3 & 14 & 48.3 & 77 & 321 & 176 \\ 
 & 6 & 1 & 6 & 18 & 51.8 & 52 & 283 & 31 \\ 
 & 7 & 1 & 3 & 27 & 48.5 & 71.5 & 396 & 290 \\ 
 & 8 & 1 & 5 & 21 & 47.8 & 65 & 209 & 210 \\ 
 & 9 & 2 & 4 & 16 & 41.3 & 54.2 & 282 & 74 \\ 
\hline
Thesaurus & 0 & 10 & 22 & 34 & 28.7 & 38 & 42 & 3 \\ 
 & 1 & 2 & 7 & 9 & 11.9 & 14 & 153 & 173 \\ 
 & 2 & 2 & 5 & 6 & 8.9 & 10 & 54 & 209 \\ 
 & 3 & 2 & 6 & 10 & 13.5 & 16 & 83 & 293 \\ 
 & 4 & 2 & 4 & 6 & 8.5 & 8.5 & 38 & 19 \\ 
 & 5 & 2 & 4 & 7 & 9.2 & 12.2 & 27 & 48 \\ 
 & 6 & 3 & 5.8 & 10.5 & 12.1 & 17.2 & 32 & 36 \\ 
 & 7 & 2 & 7 & 11 & 15.4 & 18 & 162 & 222 \\ 
 & 8 & 4 & 11 & 19 & 27.1 & 34 & 232 & 319 \\ 
 & 9 & 2 & 4 & 5 & 7.7 & 7.2 & 39 & 20 \\ 
\hline
Hand & 1 & 12 & 156 & 202 & 230.9 & 263 & 1,279 & 577 \\ 
 & 2 & 11 & 151 & 217 & 255.5 & 289 & 2,836 & 3,113 \\ 
 & 3 & 4 & 128 & 185 & 221.0 & 259 & 2,792 & 1,346 \\ 
 & 4 & 6 & 101 & 149 & 191.6 & 227 & 2,795 & 821 \\ 
 & 5 & 8 & 100 & 144 & 165.9 & 218 & 2,848 & 1,244 \\ 
 & 6 & 15 & 68 & 90.5 & 119.7 & 163 & 338 & 18 \\ 
 & 7 & 6 & 85.2 & 131 & 163.5 & 205 & 1,205 & 1,486 \\ 
 & 8 & 6 & 76 & 134 & 156.6 & 197 & 1,459 & 713 \\ 
 & 9 & 8 & 54 & 111 & 134.7 & 167.5 & 1,686 & 674 \\ 
\hline
Hand 1k & 1 & 78 & 108.5 & 145 & 153.9 & 165 & 495 & 59 \\ 
 & 2 & 31 & 105.8 & 137 & 152.8 & 183 & 657 & 748 \\ 
 & 3 & 48 & 92.5 & 113 & 121.0 & 137 & 326 & 108 \\ 
 & 4 & 22 & 52.8 & 75 & 96.4 & 106.8 & 400 & 22 \\ 
 & 5 & 39 & 61.2 & 78 & 84.8 & 95.2 & 192 & 30 \\ 
 & 7 & 26 & 51 & 77 & 85.0 & 109 & 187 & 47 \\ 
 & 8 & 23 & 53 & 66 & 74.9 & 95.5 & 160 & 15 \\ 
 & 9 & 53 & 62.2 & 69.5 & 76.5 & 95.5 & 103 & 6 \\ 
\hline
INFO+ & 1 & 16 & 97 & 171 & 237.0 & 308.5 & 1,260 & 91 \\ 
 & 2 & 17 & 95 & 183.5 & 250.3 & 329 & 2,107 & 1,344 \\ 
 & 3 & 13 & 85 & 173 & 227.8 & 307 & 1,374 & 1,533 \\ 
 & 4 & 14 & 73.8 & 141 & 197.0 & 273 & 1,245 & 336 \\ 
 & 5 & 15 & 82 & 151 & 210.5 & 288 & 1,355 & 714 \\ 
 & 6 & 14 & 83 & 181 & 249.3 & 358.8 & 1,229 & 308 \\ 
 & 7 & 8 & 78 & 144.5 & 225.4 & 289 & 1,924 & 1,540 \\ 
 & 8 & 16 & 82 & 138.5 & 227.7 & 285 & 1,318 & 1,106 \\ 
 & 9 & 22 & 75.5 & 110 & 176.3 & 209.5 & 988 & 35 \\ 
\hline
KPRM & 1 & 84 & 232.2 & 304.5 & 316.0 & 380.5 & 771 & 562 \\ 
 & 2 & 62 & 147 & 189 & 208.1 & 252 & 852 & 1,924 \\ 
 & 3 & 55 & 181 & 221 & 233.2 & 274 & 703 & 411 \\ 
 & 4 & 82 & 126.2 & 170.5 & 178.6 & 230 & 300 & 44 \\ 
\hline
Official & 0 & 4 & 18.2 & 40 & 59.8 & 77.2 & 170 & 12 \\ 
 & 1 & 1 & 6 & 28 & 70.0 & 110 & 410 & 708 \\ 
 & 2 & 1 & 5 & 37 & 67.8 & 120.8 & 508 & 2,502 \\ 
 & 3 & 1 & 4 & 29 & 59.9 & 95 & 574 & 1,961 \\ 
 & 4 & 1 & 3 & 29 & 60.4 & 117 & 287 & 257 \\ 
 & 5 & 1 & 4 & 27 & 58.3 & 101 & 307 & 521 \\ 
 & 6 & 1 & 4 & 21 & 55.4 & 107.5 & 275 & 192 \\ 
 & 7 & 1 & 4 & 32 & 60.5 & 114 & 327 & 1,466 \\ 
 & 8 & 1 & 6 & 32 & 59.5 & 104 & 340 & 1,155 \\ 
 & 9 & 1 & 4 & 17 & 39.3 & 58.8 & 396 & 426 \\ 

\end{longtable}
\end{spacing}


\end{document}